\newcommand*{\ATLASLATEXPATH}{latex/}
\pdfoutput=1 
\documentclass[UKenglish,texlive=2016,txfonts,cernpreprint]{\ATLASLATEXPATH atlasdoc}
\usepackage{\ATLASLATEXPATH atlaspackage}
\usepackage{\ATLASLATEXPATH atlasbiblatex}
\usepackage{\ATLASLATEXPATH atlasphysics}
\graphicspath{{logos/}{figures/}}
\usepackage{JpsiPlusMu8TeVPaper-defs}

\usepackage{lscape}
\usepackage{verbatim}
\usepackage{pdflscape}
\usepackage{graphicx}
\usepackage{amssymb}
\usepackage{array}
\usepackage{siunitx}

\AtlasTitle{Measurement of \Bhadron pair production with the ATLAS detector in proton-proton collisions at $\sqrt{s}=8$~\TeV}

\author{The ATLAS Collaboration}

\AtlasJournal{JHEP}
\PreprintIdNumber{CERN-EP-2017-057}
\AtlasDate{<Date>}
\AtlasJournalRef{JHEP 11 (2017) 62}
\AtlasDOI{10.1007/JHEP11(2017)062}
\arXivId{1705.03374}

\AtlasAbstract{%
A measurement of \Bhadron pair production is presented, based on a
data set corresponding to an integrated luminosity of 11.4~fb$^{-1}$ of
proton--proton collisions recorded at $\sqrt{s}=8$~TeV with the ATLAS
detector at the LHC. Events are selected in which a \Bhadron is
reconstructed in a decay channel containing
$\Jpsi \rightarrow \mu\mu$, and a second \Bhadron is
reconstructed in a decay channel containing a muon. Results are
presented in a fiducial volume defined by kinematic requirements on
three muons based on those used in the analysis. The fiducial cross
section is measured to be $17.7 \pm 0.1 ($stat.$) \pm 2.0
($syst.$)$~nb. A number of normalised differential cross sections are
also measured, and compared to predictions from
the \pyegt, \hwpp, \mgamc{}+{}\pyegt\ and \SHERPA event generators,
providing new constraints on heavy flavour production.}

\hypersetup{pdftitle={ATLAS document},pdfauthor={The ATLAS Collaboration}}

\begin{document}

\newcommand*{\numR}[2]{\num[round-precision=#2]{#1}}
\newcommand*{\numRF}[2]{\num[round-mode=figures,round-precision=#2]{#1}}
\newcommand*{\numRP}[2]{\num[round-mode=places, round-precision=#2]{#1}}
\newcommand{\numpmerr}[3]{\ensuremath{^{\num[round-precision=#3]{#1}}_{\num[round-precision=#3]{#2}}}}
\newcommand{\numpmRF}[3]{\ensuremath{^{\num[round-mode=figures,round-precision=#3]{#1}}_{\num[round-mode=figures,round-precision=#3]{#2}}}}
\newcommand{\numpmRP}[3]{\ensuremath{^{\num[round-mode=places, round-precision=#3]{#1}}_{\num[round-mode=places, round-precision=#3]{#2}}}}

\maketitle

\section{Introduction}
\label{sec:intro}
 
The production of $b$- and $c$-quarks (heavy flavour) in
proton--proton collisions provides a fruitful testing ground for the
predictions of quantum chromodynamics (QCD). The mass of the $b$-quark
introduces an additional scale, and the typical factorisation of QCD
calculations into parton distribution functions, hard matrix element
and softer parton shower components allows this mass to be
introduced at different stages. Furthermore, there are several
possible schemes for the inclusion of the heavy-quark masses at these
various stages and, given this freedom, the optimal settings must be
determined by comparisons to data.

Several techniques have previously been used to measure heavy-flavour
production. Some measurements rely on identifying hadronic
jets containing a \Bhadron\footnote{In this paper ``\Bhadron'' and
  ``\Chadron'' are used as a shorthand for any weakly decaying
  hadron containing a valence $b$-\ or \bbar-quark, or a valence $c$-\ or
  \cbar-quark, respectively.} based on the presence of a decay vertex
or charged-particle tracks displaced from the primary interaction vertex due to
the lifetime of the \Bhadron. These measurements have been performed
inclusively~\cite{ATLAS:2011ac, Chatrchyan:2012dk, Aad:2012ma, Aaboud:2016jed} and in
the associated production of a vector boson ($V$) and jets~\cite{Aad:2014dvb,
  Chatrchyan:2014dha, Aad:2013vka, Chatrchyan:2013uza, Abazov:2014fka,
  Abazov:2013uza, Abazov:2015aha, PhysRevLett.111.042003,
  PhysRevLett.104.131801, Aaltonen:2008mt}.  Other classes of
measurements identify semileptonic \Bhadron decays through the
presence of an electron or muon~\cite{Chatrchyan:2012hw, Aad:2011rr,
  Abbott:1999se, Aaltonen:2007zza, Abbott:1999wu,
  PhysRevLett.74.3548}.  Finally, several measurements identify
displaced decay vertices without any jet
requirements~\cite{Khachatryan:2011wq, Chatrchyan:2013zja}, or select
specific decay channels such as $B^{\pm}\rightarrow\Jpsi+X$~\cite{Aad:2014kba, Aad:2012jga, Aad:2011sp, Aad:2015duc,
  Khachatryan:2010yr, Acosta:2004yw, ATLAS:2013cia}.

These measurements have highlighted disagreements both between different theoretical
predictions and between those predictions and the data, motivating further study. Specifically, the
region of small-angle $b\bar{b}$ pair production is particularly
sensitive to the details of the various calculations, but so far
remains only loosely constrained by the experimental
data. Understanding this region, and heavy-flavour production more
generally, is also increasingly relevant at the LHC where, for
example, searches for the Higgs boson produced in association with a
vector boson ($VH$) and decaying to a \bbbar pair rely extensively on
the modelling of one of the main backgrounds: QCD production of \bbbar
pairs in association with vector bosons. These searches are most
sensitive when the Higgs boson has significant momentum and the
resulting $b$-quarks are produced at small opening angles, so
constraining the theoretical predictions for the background in this
region is an important input to these searches.
 
This paper describes a new measurement of the production of two
\Bhadrons, where one \Bhadron decays to $J/\psi(\rightarrow\mu\mu)+X$
and the other to $\mu+Y$, resulting in three muons in the final
state. The signal definition includes \Jpsi produced from feed-down
from excited charmonium states, as well as muons produced in
semileptonic cascade decays (the decay of a \Chadron produced in the
decay of a \Bhadron). To probe \Bhadron production, several
differential cross sections are measured, based on the kinematics of
the \Jpsi (reconstructed from two muons) and the third muon. The variables
considered are:
\begin{itemize}
\item the azimuthal separation between the \Jpsi and third muon,
$\Delta\phi(\Jpsi,\mu)$;\footnote{ATLAS uses a right-handed coordinate system
with its origin at the nominal interaction point (IP) in the centre of
the detector and the $z$-axis along the beam pipe. The $x$-axis points
from the IP to the centre of the LHC ring, and the $y$-axis points
upward. Cylindrical coordinates $(r,\phi)$ are used in the transverse
plane, with $\phi$ being the azimuthal angle around the $z$-axis.
Pseudorapidity is defined in terms of the polar angle $\theta$ as
$\eta=-\ln\tan(\theta/2)$. Transverse momentum (\pt) is defined as the projection of the momentum in the
transverse plane.}
\item the transverse momentum of the three-muon system, $p_{T}(\Jpsi, \mu)$;
\item the separation between the \Jpsi and the third muon in the
  azimuth--rapidity plane, $\DRy(\Jpsi,\mu)$,\footnote{$\DRy(A,B)$ is
    defined as the separation between particles $A$ and $B$ in the
    azimuth--rapidity ($y$) plane, $\DRy(A,B) =
    \sqrt{\Delta\phi(A,B)^2 + \Delta y(A,B)^2}$.} measured inclusively
  and split into $\pT(\Jpsi, \mu)<20$~GeV (low \pt) and $\pT(\Jpsi,
  \mu)\geq20$~GeV (high \pt) regions;
\item the separation in rapidity between the \Jpsi and the third muon,
  $\Delta y(\Jpsi,\mu)$;
\item the magnitude of the average rapidity of the \Jpsi and the third muon, $y_{\textrm{boost}}$;
\item the mass of the three-muon system, $m(\Jpsi, \mu)$;
\item the ratio of the transverse momentum of the three-muon system to the invariant mass of
the three-muon system, $\pT/m$, and its inverse,
$m/\pT$. 
\end{itemize}
Results are presented as the total cross section and normalised differential cross
sections, all within a fiducial region defined in Section~\ref{sec:fiducial}.

\section{ATLAS detector}
\label{sec:detector}

The ATLAS detector~\cite{PERF-2007-01} consists of an inner tracking
system, referred to as the inner detector, surrounded by a
superconducting solenoid, electromagnetic and hadronic calorimeters
and an external muon spectrometer. Charged-particle tracks in the
pseudorapidity range $|\eta| < 2.5$ are reconstructed with the inner
detector, which is immersed in a 2~T axial field provided by the
solenoid. The inner detector consists of pixel and
microstrip semiconductor detectors, as well as a straw-tube transition
radiation tracker. The solenoid is surrounded by sampling
calorimeters, which span the pseudorapidity range up to $|\eta| =
4.9$. High-granularity liquid-argon (LAr) electromagnetic calorimeters
are present up to $|\eta| = 3.2$.  Hadronic calorimeters with
scintillator tiles as active material cover $|\eta| < 1.74$, while LAr
technology is used for hadronic calorimetry from $|\eta| = 1.5$ to
$|\eta| = 4.9$. Outside the calorimeter system, air-core toroids
provide a magnetic field for the muon spectrometer. Three stations of
precision drift tubes and cathode-strip chambers provide measurements
of muon tracks in the region $|\eta| < 2.7$. Resistive-plate and
thin-gap chambers provide muon triggering capability up to $|\eta| =
2.4$.

\section{Data set, event selection and simulation}
\label{sec:dataset}

\subsection{Data set and event selection}
This analysis uses proton--proton collision data recorded by the ATLAS
experiment during 2012 at a centre-of-mass energy $\sqrt{s} =
8$~TeV. All events considered were recorded while the detector and
trigger systems were functional and satisfied data quality
requirements. Events are selected using a dimuon trigger where the
muons are required to have opposite charge, be consistent with
originating from the same production vertex, have $\pt(\mu)>4$~GeV and
$|\eta(\mu)|<2.4$ and satisfy a loose dimuon mass selection,
$2.5<m(\mu^+,\mu^-)<4.3$~GeV. The trigger was prescaled for various
periods throughout 2012, and the integrated luminosity of the
resulting data set is $11.4$~fb$^{-1}$.

Muon candidates are reconstructed by combining inner detector tracks
with tracks in the muon spectrometer. Candidates are required to have
$|\eta| < 2.5$ and $\pt > 6$~GeV. Additional requirements are placed
on the track quality, requiring a minimum number of hits in the
different layers of each of the inner detector subcomponents. The \Jpsi
candidates are formed by selecting pairs of muons with opposite charge that
have been identified to originate from a common vertex. These muons 
must also have $|\eta| < 2.3$ to lie within the trigger acceptance, and
are required to match the direction of the corresponding trigger-level
candidates. The invariant mass of the \Jpsi candidate is required to be
in a window around the Particle Data Group (PDG) average of the \Jpsi
mass of 3.097~GeV~\cite{Agashe:2014kda}, $m(\Jpsi_{\textrm{PDG}})$, specifically in the range
2.6--3.5~GeV. If more than one dimuon \Jpsi candidate satisfies the
requirements above, the one with a reconstructed mass closest to the
PDG mass is chosen.

A third-muon candidate is also required in the event. If there are
more than three muons in an event the highest-\pT muon that is not
used in the \Jpsi candidate construction is defined to be the third
muon.

\subsection{Simulation}

Inclusive \Bhadron pair production is simulated with the
\pyegt.186~\cite{Sjostrand:2007gs} Monte Carlo event generator, based
on a $2 \to 2$ matrix element calculation matched to a parton shower.
The \textsc{CTEQ6L1}~\cite{Pumplin:2002vw} parton distribution
function (PDF) set is used, along with the
\textsc{AU2}~\cite{ATL-PHYS-PUB-2012-003} tuned parameter
settings. The $b$-quarks are treated as massless in the PDF set and
the matrix element calculation, but the mass is reinstated during the
parton shower. Events are filtered based on the presence of
$J/\psi(\rightarrow \mu\mu)$ produced in the decay of a \Bhadron,
requiring two muons with $\pt > 6$~GeV.\footnote{All \Jpsi are decayed
  to $\mu\mu$. In events where two \Jpsi are present, the event is
  weighted by BR($\Jpsi\rightarrow\mu\mu$) to correct for this.}  The
\MCd collisions are overlaid with additional simulated minimum-bias
collisions, to emulate the effect of multiple proton--proton
interactions occurring during the same (in-time) or a nearby
(out-of-time) bunch crossing, an effect called ``pile-up''. These
additional collisions are produced using \pyegt with the
\textsc{A2}~\cite{ATL-PHYS-PUB-2011-014} tuned parameter settings and
the \textsc{MSTW2008}~\cite{Martin:2009iq} PDF set.  The \MCd events
are then passed through a GEANT4~\cite{Agostinelli:2002hh} simulation
of the full ATLAS detector~\cite{Aad:2010ah}, and reconstructed using
the same software as the real data.

In addition, an inclusive $pp\rightarrow b\bar{b}$ sample is
simulated with the \hwpp~\cite{Bahr:2008pv} event generator using the
\textsc{CTEQ6L1} PDF set with \textsc{UE-EE5}~\cite{Gieseke:2012ft}
parameter settings tuned for the underlying-event modelling.  Again,
this prediction includes a $2 \to 2$ matrix element calculation
matched to a parton shower, but in this case, the $b$-quarks are
considered massive in both the matrix element calculation and parton
shower.

For both predictions, all $2 \to 2$ QCD processes are included and
$b$-quarks can be produced either in the matrix element or in the
subsequent parton shower phase.

\subsection{Fiducial volume definition}
\label{sec:fiducial}
Cross sections are defined in terms of observables at the particle
level, defined in terms of particles with an average lifetime greater
than 10~mm/$c$. The signal is defined by two final-state muons that
originated from the decay of a \Jpsi, which itself is a descendant of
a \Bhadron (including feed-down decays), and a third muon which
originated from a different \Bhadron (including cascade decays).
Muons are ``dressed'' by adding the four-momentum of photons
(excluding photons produced in the decays of hadrons) that are close
to the muon, defined by $\DReta(\mu,\gamma) <
0.1$,\footnote{$\DReta(A,B)$ is defined as the separation in the
  azimuth--pseudorapidity plane, $\DReta(A,B) =
  \sqrt{\Delta\phi(A,B)^2 + \Delta\eta(A,B)^2}$.} to the muon
four-momentum. The resulting dressed muons are required to have $\pt >
6$~GeV. The two muons from the \Jpsi are required to have $|\eta|<2.3$
and the third muon is required to have $|\eta|<2.5$, matching the
trigger and reconstruction acceptances respectively.

\section{Muon trigger and reconstruction efficiencies}
\label{sec:muontrigreco}
The reconstructed \Jpsi and third muon candidates are taken from muons
satisfying the selection criteria previously described. To extract the
fiducial cross sections, the data must be corrected to account for
inefficiencies in the selection of events containing
muons. Corrections for the muon trigger and reconstruction
efficiencies are obtained from data-driven techniques and applied to
the observed events via an event weight.

The trigger efficiency correction~\cite{Aad:2012dlq} is factored
into several components: 
\begin{itemize}
\item The single-muon trigger efficiency is calculated with a
  tag-and-probe method using \Jpsi candidate data. The efficiency is
  dependent on the kinematics of the muon and is parameterised as a
  function of \pt and $q \times \eta$, where $q$ is the reconstructed
  muon charge. A correction derived from \MCn is required to remove
  bias arising from the difference between the single-muon triggers
  used for the tag-and-probe method (which require $\pt(\mu)>18$~GeV), and
  the dimuon trigger used in the analysis (which requires
  $\pt(\mu)>4$~GeV for both muons).
\item The dimuon trigger efficiency correction is calculated with data
  in two parts, a vertex-finding and opposite-charge correction, and a
  correction for the spatial overlap of muons in the trigger system,
  which typically results in a reduction in efficiency when
  \DReta$(\mu,\mu)<0.2$. The spatial overlap correction is calculated
  as a function of the angular separation between the two muons
  calculated in three separate dimuon rapidity intervals.
\item A correction taken from \MCn is applied to account for trigger
  inefficiencies in events with three muons where a pair of muons
  falls in the same trigger-level object, also typically for
  \DReta$(\mu,\mu)<0.2$.  This correction is derived by fitting a
  linear function to the ratio of the number of \Bhadron pair
  production events accepted by the trigger to all signal events, as a
  function of the separation of the third muon from the closest muon
  in the \Jpsi candidate.
\end{itemize}
The per-event trigger efficiency is given by the product of each
single-muon efficiency, the additional dimuon trigger efficiency
corrections and the three-muon efficiency correction.

The data are also corrected for the muon reconstruction
efficiency~\cite{Aad:2014rra} which is applied per muon for each of
the three muons in the event. The efficiency for a single muon is
calculated in two parts: the efficiency for a muon track to be
reconstructed in the inner detector is parameterised as a function of
muon \pt and $\eta$; then the efficiency for reconstructing a muon
given that an inner detector track has already been reconstructed is
calculated using a tag-and-probe method using
$Z\rightarrow\mu^{+}\mu^{-}$\ and $J/\psi\rightarrow\mu^{+}\mu^{-}$
data. The correction is derived as a function of muon \pt and $q
\times \eta$.

\section{Signal extraction}
\label{sec:signal}

While the event selection described in Section~\ref{sec:dataset}
provides a high-purity \Jpsi$+\mu$ sample, the signal of interest in
this analysis must still be extracted from the data. This is done in
the following stages:
\begin{enumerate} 
\item The yield of \Jpsi resulting from the decay of a \Bhadron (the
  signal) is extracted using a simultaneous fit to the distributions
  of dimuon mass and pseudo-proper decay time, $\tau$, defined as
  \begin{equation}
    \tau = L_{xy} \times m(\Jpsi_{\textrm{PDG}}) / \pT(\mu^+\mu^-),
  \end{equation}
  where $L_{xy}$ is the transverse distance between the primary vertex
  and the dimuon vertex positively (negatively) signed for a vertex
  with momentum vector pointing away from (towards) the primary
  vertex. The primary vertex is defined as the vertex formed from at
  least two tracks, each with $\pt > 400$~MeV, that has the largest
  summed track $\pt^2$ in the event. The quantity
  $\pT(\mu^+\mu^-)$\ is the transverse momentum of the dimuon
  system. This fit is described in Section~\ref{sec:fittingjpsi}, and
  the resulting determination of \Jpsi background contributions is
  used as an input to the next step.

\item Next, the yield of events containing an additional (i.e. third)
  muon resulting from the decay of a \Bhadron is determined. First,
  the proportion of signal relative to background is enhanced by
  requiring $\tau > 0.25$~mm/c. The remaining contribution from
  certain backgrounds is determined from the \Jpsi fit, and other
  backgrounds are determined using a simultaneous fit to the
  transverse impact parameter significance, $\dzerosig$, and the
  output of a boosted decision tree (BDT) trained to separate signal
  muons from misreconstructed muons, as described in
  Section~\ref{sec:fitThirdMu}. The transverse impact parameter
  significance is defined as
  \begin{equation} 
    \dzerosig \equiv \dzero/\sigma_{\dzero},
  \end{equation}
  where the transverse impact parameter, \dzero, is the distance of
  closest approach of the track to the primary vertex point in the
  $r$--$\phi$ projection, with the \dzero sign given by the sign of
  the angular momentum of the track around the beam evaluated at the
  point of closest approach; and $\sigma_{\dzero}$ is the (unsigned)
  uncertainty in the reconstructed \dzero.
\item Some remaining irreducible sources of background are then
  subtracted from the fitted yields, as described in
  Section~\ref{sec:irreducibles}.
\item Having determined the yield of \Jpsi and third muons, this is
  corrected for the effect of the $\tau$ requirement ($\tau >
  0.25$~mm/c), as described in Section~\ref{sec:extrapolation}.
\item Finally, the effects of detector resolution are corrected for, as
  described in Section~\ref{sec:unfold}, determining the measured cross
  section for the signal.
\end{enumerate} 

This process is repeated for each kinematic bin of each differential
cross section, resulting in the cross section in that bin. The width
of these bins was chosen in order to retain the maximum information
about the shape of each differential cross section, while maintaining
a sufficient number of events in each bin to minimise the
uncertainties. The data used in all fits are already corrected by the
muon trigger and reconstruction efficiencies previously described.

\subsection{Extraction of the \Jpsi composition}
\label{sec:fittingjpsi}

To identify the \Jpsi signal and background contributions to the data,
a two-dimensional unbinned extended maximum-likelihood fit to the
\Jpsi candidate mass and pseudo-proper decay time is performed. There
are two components of the \Jpsi candidate dimuon mass spectrum: the
real \Jpsi contribution, which is peaked at the \Jpsi mass; and the
fake \Jpsi background contribution, which forms a continuum under the
\Jpsi peak. The pseudo-proper decay time distribution for a real \Jpsi
has two components. The first is for a \Jpsi from direct strong
production in the hard scatter; this component is peaked at low $\tau$
and is referred to as \emph{prompt}. The second corresponds to the
\Jpsi signal from \Bhadron decays that has, on average, larger values
of $\tau$ due to the lifetime of the \Bhadron, and is referred to as
\emph{non-prompt}. Both prompt and non-prompt \Jpsi include feed-down
from excited charmonium states produced in either the hard scatter or
the decay of a \Bhadron.  Similarly, there are three components of the
fake \Jpsi background included in the fit, to account for different
contributions to the pseudo-proper decay time (see below). The
pseudo-proper decay time probability density function (p.d.f)
components are convolved with a detector resolution function, modelled
by a double Gaussian distribution centred at $\tau = 0$~mm/c, with the
widths of those Gaussian functions and their relative normalisation
allowed to float freely in the fit.

In order to extract the number of non-prompt \Jpsi, a fit model with
five functional forms is used, based on the model used in
Ref.~\cite{Aad:2015duc}. The forms are:
\begin{itemize}
\item \textit{Non-prompt \Jpsi:} The dimuon invariant mass is modelled
  using the sum of a Crystal
  Ball~\cite{Oreglia:1980cs,Gaiser:1982yw,Skwarnicki:1986xj} and a
  Gaussian function. The $\tau$ distribution is modelled using a
  single-sided exponential decay function, convolved with the
  resolution function.

\item \textit{Prompt \Jpsi:} The dimuon invariant mass is modelled
  using the same Crystal Ball and Gaussian function used for the
  non-prompt \Jpsi. The $\tau$ distribution is modelled using a delta
  function at $\tau=0$~mm/c, convolved with the resolution function.

\item \textit{Prompt fake \Jpsi background:} The dimuon invariant
  mass is modelled using a constant distribution
  (fits using a first order polynomial yield slopes consistent with zero
for this component). The $\tau$ distribution is modelled using a delta
  function at $\tau=0$~mm/c, convolved with the resolution function.

\item \textit{Single-sided fake \Jpsi background:} The dimuon
  invariant mass distribution is modelled by a negative-slope
  exponential function. The $\tau$ distribution is modelled using a
  single-sided exponential decay function, convolved with the
  resolution function.

\item \textit{Double-sided fake \Jpsi background:} The dimuon
  invariant mass distribution is modelled by an exponential decay
  function. The $\tau$ distribution is modelled using a double-sided
  exponential decay function, convolved with the resolution function.
\end{itemize}

In the mass fit, the Gaussian and Crystal Ball functions share the
same mean value. The Crystal Ball $n$ and $\alpha$ parameters are
fixed to values derived from an inclusive \Jpsi fit, but the width is
allowed to float. The mean values of the decay time model's Gaussian
function and double-sided exponential functions are fixed at zero. All
other fit parameters are allowed to float.

\begin{figure}[t!]
   \centering
   \includegraphics[width=0.99\textwidth]{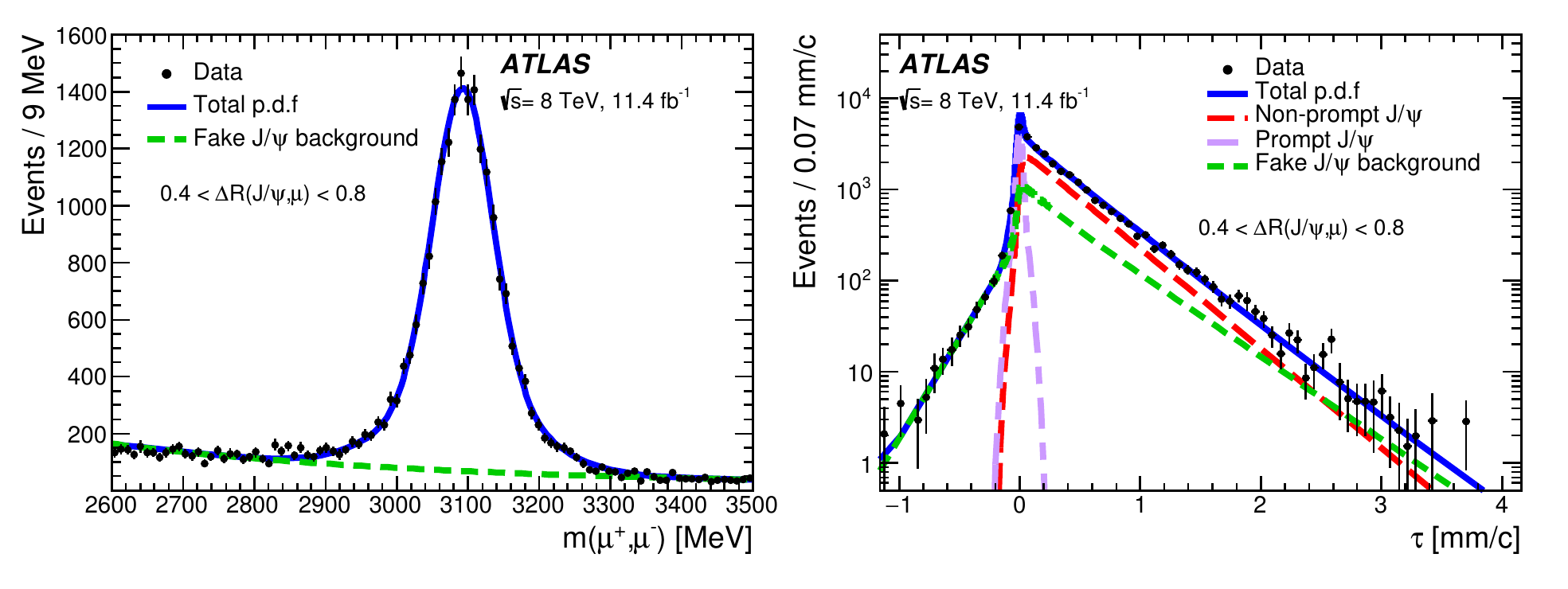}
 \caption{The two-dimensional dilepton mass--pseudo-proper decay time
   fit for a single differential observable bin: 0.4 <
   $\DRy(\Jpsi,\mu)$ < 0.8.  The points with error bars are data,
   plotted at bin centres. The solid line is the projection of the
   unbinned maximum-likelihood fit to the data. The dilepton invariant
   mass projection (left) shows the total fake \Jpsi background
   component (dashed) in addition to the total p.d.f. The
   pseudo-proper decay time projection (right) also shows the prompt
   and non-prompt signal \Jpsi as dashed lines.\label{fig:Jpsifit}}
\end{figure}

The functional forms are combined into the fit model used in the
unbinned extended maximum likelihood fit to the data. The fit for an
example kinematic bin is shown in Figure~\ref{fig:Jpsifit}, where the
prompt, single-sided and double-sided fake \Jpsi backgrounds are
combined for clarity.

The stability and performance of the fit are checked by a closure test
on \MCd samples. Pseudo-data sets are produced by combining different
numbers of prompt and non-prompt \Jpsi events. The two-dimensional
\Jpsi model is then fitted to each pseudo-data set and the fractions
of prompt and non-prompt \Jpsi events is extracted from the fit. The
fractional difference between the input and fitted number of prompt
and non-prompt \Jpsi events is compared, demonstrating that the fit is
performing very well with deviations from the input composition
consistently below 2\%, which is comparable to the statistical
uncertainty in the fit.

\subsection{Extraction of the non-prompt muon signal}
\label{sec:fitThirdMu}

Having determined the \Jpsi contributions, the next step is to extract
the signal yield: a non-prompt muon from the same hard scatter as a
non-prompt \Jpsi. To do this, a second two-dimensional
maximum-likelihood fit is performed, in this case using two
observables that allow separation of the signal muons from background:
the transverse impact parameter significance, and a BDT trained to
separate signal muons from instrumental backgrounds.

The mechanisms of prompt and non-prompt \Jpsi production differ
significantly, resulting in very different background contributions to
the third-muon sample in each case. This leads to difficulties in
fitting the third-muon distributions across the entire \Jpsi $\tau$
range. To increase the signal muon purity and improve the third-muon
fit performance, the selected \Jpsi in each event is first required to
have $\tau > 0.25$~mm/c, removing all of the prompt \Jpsi
candidates. In addition, to reduce the \Jpsi candidates arising from
the continuum background, a tighter invariant mass requirement of
$2.95 < m(\mu^+,\mu^-) < 3.25$~GeV is applied. To account for the
signal efficiency loss from these criteria, a correction is made once
the signal yield is extracted from the fits.

There are several contributions to the third-muon background that are
then considered: prompt muons produced at the primary vertex, muons
produced in the decays of charged pions or kaons, third muons in
events where the \Jpsi candidate is not a real \Jpsi but from the
continuum background, and events where the \Jpsi and third muon are
produced from different hard scatters in the same bunch crossing
(referred to as the pile-up background).

There are two sources of \emph{fake} muon background.
\emph{Decay-in-flight} (DIF) muons are the result of the decay of a
charged pion or kaon. The small mass difference between the hadrons
and resulting muons can result in only a small deflection of the track
at the decay vertex. Therefore, the track of both the hadron and
resulting muon can be reconstructed as coming from a single particle
and a muon candidate may be identified. Muon candidates can also be
reconstructed when charged hadrons leave tracks in the inner detector
and charged particles from the shower in the hadronic calorimeter
leave tracks in the muon spectrometer, referred to as \emph{hadronic
  shower leakage} muons. Both the DIF and hadronic shower leakage
sources of fake muons contribute significantly at low angular
separations between the \Jpsi and third muon, as they can arise
through decays such as $B^{\pm}\rightarrow\Jpsi+K^{\pm}$.

To discriminate between the signal and fake muons, a BDT is
constructed from kinematic variables that are sensitive to the
production mechanism of the muons. These variables are:

\begin{itemize}
\item \emph{Track deflection significance}: This parameter is the
  maximum value of the significance of the difference in track
  curvature calculated upstream and downstream of a point somewhere
  along the track reconstructed in the inner detector.  DIF muons
  originating from a point inside the inner detector typically have
  higher values of track deflection significance than the signal
  muons.
\item \emph{Track deflection neighbour significance}: Computed
  considering track segments, between adjacent hits, along the inner
  detector track. The largest value over the whole track of the
  significance of the angular difference between adjacent track
  segments is taken. This variable quantifies the significance of a
  deflection along a muon track; DIF muons originating from a point
  inside the inner detector populate larger values.
\item \emph{Momentum balance significance}: The significance of the
  difference between the track momenta reconstructed in the inner
  detector and in the muon spectrometer. If a pion or kaon decays
  outside the inner detector, the inner detector track may be matched
  to a lower-momentum muon spectrometer track produced by the
  resulting muon. The imbalance between the two track momenta is
  higher for these DIF muons than for inner detector and muon
  spectrometer tracks produced by a single muon. This variable also
  offers discrimination between signal and hadronic shower leakage
  muons.

\item \emph{Absolute pseudorapidity}, $|\eta|$: Muon candidates
  produced in the background processes are more likely to be produced
  at high absolute pseudorapidities.
\end{itemize}

\begin{figure}[htbp]
  \centering
  \includegraphics[width=0.49\textwidth]{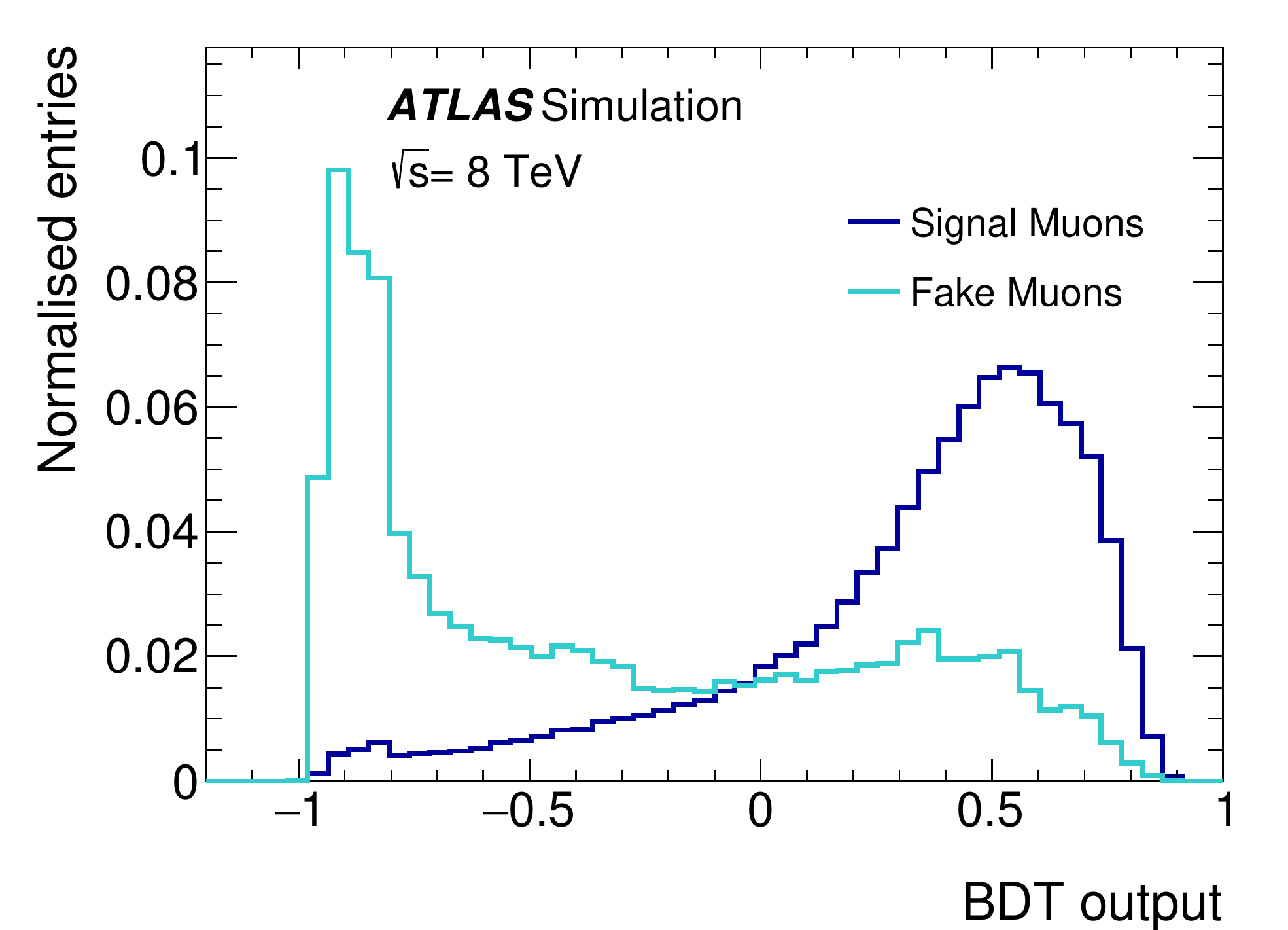}
  \includegraphics[width=0.49\textwidth]{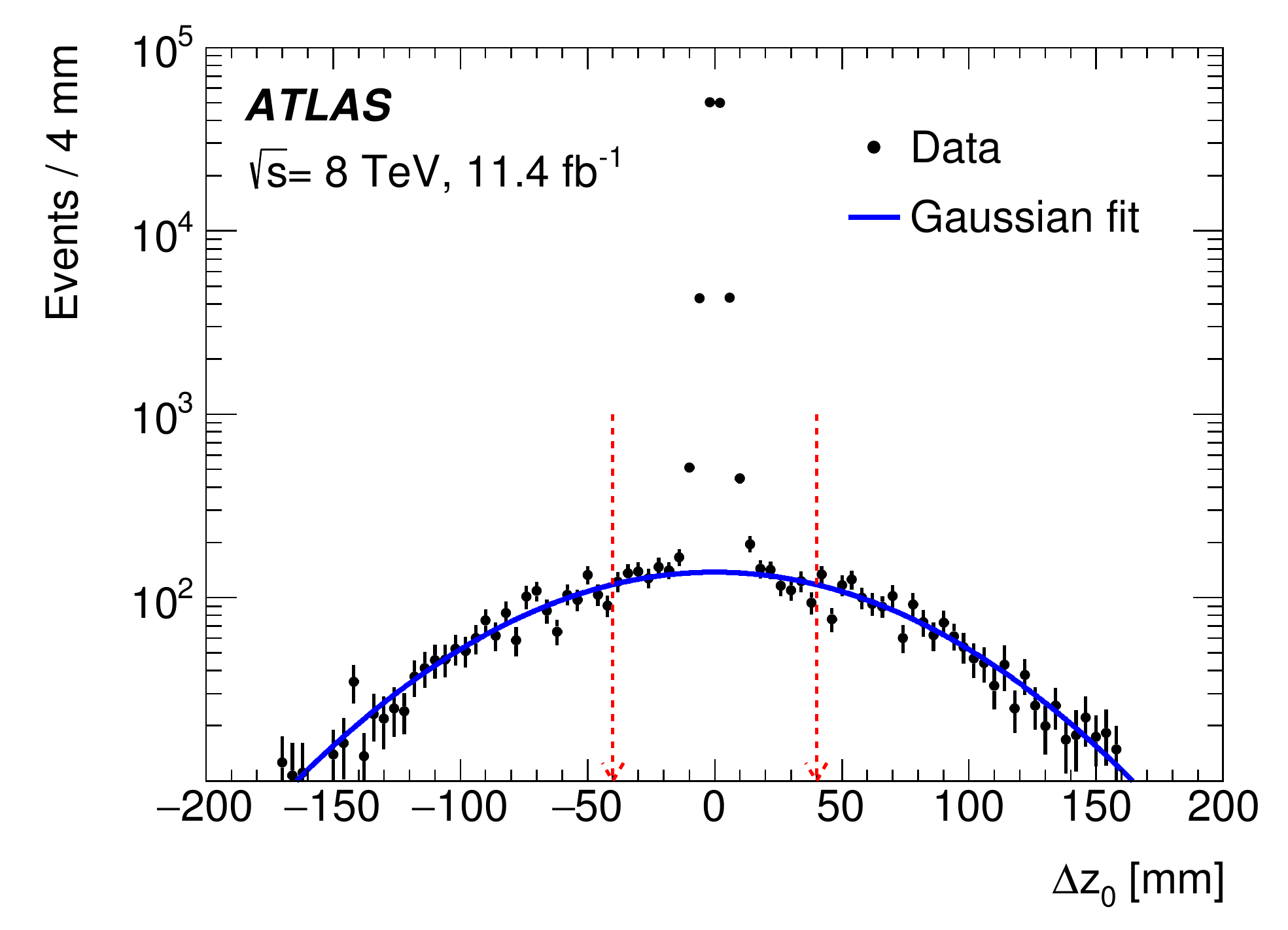}
  \caption{Left: Distribution of the BDT output for signal muons (dark
    blue) and background fake muons (light blue) taken from
    simulation. Right: Data $\Delta z_{0}$ distribution for the
    inclusive data set including the Gaussian pile-up background
    fit. The Gaussian function is fitted using the data outside the
    region marked by the dashed arrows. \label{fig:ThirdmuBDTPU}}

\end{figure}

The BDT is developed in the TMVA framework~\cite{Hocker:2007ht}. The
\pyegt \MCn is separated into two independent samples; one is used for
training the BDT and the other is used to validate the
performance. Figure~\ref{fig:ThirdmuBDTPU} (left) shows a clear
discrimination between the distribution of the BDT output for the
signal and the fake muon background.

The BDT is trained using signal muons and fake muons, both taken from
the \MCd signal sample using the list of particles produced in each
simulated collision, the Monte Carlo (MC) event record. Signal muons
are defined as reconstructed muons, $\mu_{\textrm{reco}}$, that can be
associated to a muon from the \MC event record, $\mu_{\textrm{MC}}$,
which was produced in the decay of a \Bhadron.  The association
between reconstructed and simulated muons is performed by requiring
$\DReta(\mu_{\textrm{MC}}, \mu_{\textrm{reco}})<0.02$.  Muons from DIF
are identified as reconstructed muons which fail the signal
definition, but are associated to a muon in the \MC event record which
has a charged pion or kaon as a parent. In this case, a looser
association is required due to the change in the momentum at the point
the pion/kaon decays: $\DReta(\mu_{\textrm{MC}},
\mu_{\textrm{reco}})<0.15$.  Finally, the remaining reconstructed
muons may be identified as hadronic shower leakage if they do not
match any muon from the MC event record, but do match a charged pion
or kaon. Again, a looser association of $\DReta(\pi/K_{\textrm{MC}},
\mu_{\textrm{reco}})<0.15$ is used to correctly identify all of this
background.

There is also a background of third muons in events where the \Jpsi
candidate comes from the dimuon continuum background (the ``fake
\Jpsi'' contribution introduced in Section~\ref{sec:fittingjpsi}).
Events lying in the regions outside the dimuon signal mass region
($2.95 < m(\mu^+,\mu^-) < 3.25$~GeV) but within the range $2.60 <
m(\mu^+,\mu^-) < 3.50$~GeV consist almost entirely of this fake \Jpsi
background. These events are therefore used to form the \dzerosig and
BDT templates used for the fake \Jpsi background. These templates are
normalised using the results from the two-dimensional \Jpsi fit
(described in Section~\ref{sec:fittingjpsi}) to determine the number
of fake \Jpsi events in the signal mass region.

The background from pile-up is studied using $\Delta z_0$, defined as
the difference between the reconstructed $z$-position (at their
respective points of closest approach to the beam axis) of the
third-muon track and the \Jpsi candidate muon which maximises the
value of $\Delta z_0$. Figure \ref{fig:ThirdmuBDTPU} (right) shows the
$\Delta z_{0}$ distribution for data after all event selection
criteria are applied. The distribution consists of two components: a
peaked structure centred on zero, representing events where the \Jpsi
candidate and third muon are produced in the same proton--proton
interaction, and a Gaussian-distributed component from pile-up
spanning a wide $\Delta z_0$ range.  To suppress the pile-up
background, events are required to have $|\Delta z_{0}| < 40$~mm, and
the pile-up background within this signal region is estimated by
fitting a Gaussian model to the broad $\Delta z_0$ distribution,
excluding the signal region from the fit. The integral of the Gaussian
function within $|\Delta z_{0}| < 40$~mm gives the number of residual
pile-up events in the signal region, and the shape of third muon BDT
and \dzerosig distributions for pile-up events is taken from the pure
pile-up region outside this $|\Delta z_{0}|$ range.

In summary, the two-dimensional fit of \dzerosig vs BDT for the third
muon is constructed from the following components:

\begin{itemize}
\item \emph{Signal $\mu$}: Both the BDT and \dzerosig fit templates
  are taken from the \pyegt \MCd sample, using reconstructed muons
  matched to a muon in the \MCd event record which derives from a
  \Bhadron. This component is expected to populate the high values of
  the BDT output, signifying real muons, and have a wide \dzerosig
  distribution indicating production away from the interaction
  point. The shapes of the templates are fixed but the normalisation
  floats in the fit.

\item \emph{Prompt $\mu$}: Both the BDT and \dzerosig templates are
  taken from \Jpsi muons in the inclusive $pp\rightarrow \Jpsi$ \pyegt
  \MCn, where \Jpsi production is dominated by prompt
  production. These muons are real and thus should occupy the high
  values in the BDT output distribution and will have a narrow
  \dzerosig distribution as they are produced at the interaction
  point. The shapes of the templates are fixed but the normalisation
  floats in the fit.

\item \emph{Prompt and non-prompt fake $\mu$}: Both the BDT and
  \dzerosig fake muon templates are taken from the \pyegt \MCd
  sample. The same BDT template is used for both the prompt and
  non-prompt components. The BDT shape has a large contribution at low
  values. The \dzerosig template is derived separately for the prompt
  and non-prompt components as fake muons can have both prompt as well
  as non-prompt sources. The shape of the templates are fixed but the
  normalisation floats in the fit.

\item \emph{Fake \Jpsi}: The BDT and \dzerosig templates are derived
  from data and fixed in the fit.

\item \emph{Pile-up}: The BDT and \dzerosig templates are derived from
  data and fixed in the fit.
\end{itemize}

The three \dzerosig templates derived from \MCn have a small shift in
the mean of the distribution relative to data, due to the modelling of
the beam-spot position. To correct for this, a small shift in the mean
is derived independently for each of these three templates, by
rerunning the fit once and allowing the shift to float; the shifted
mean is then fixed at the fitted value for all other fits. The shifts
for the three samples are consistent within uncertainties, and
approximately $-0.05$.

\begin{figure}[t!]
   \centering
     \includegraphics[width=0.99\textwidth]{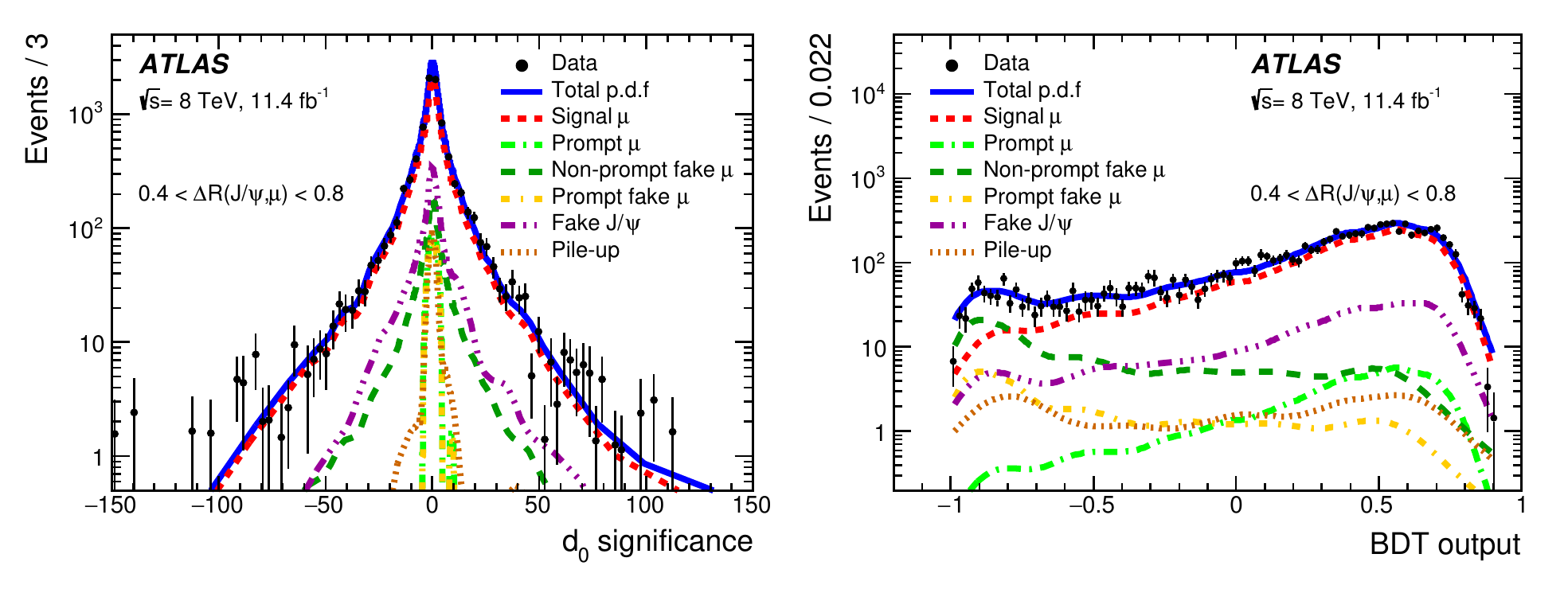}
     \caption{The two-dimensional \dzerosig{}(left)--BDT(right) fit
       for a single differential observable bin: 0.4 <
       $\DRy(\Jpsi,\mu)$ < 0.8. The points with error bars are data,
       plotted at bin centres. The solid line is the projection of the
       unbinned maximum-likelihood fit to the data. The signal $\mu$,
       prompt $\mu$, non-prompt fake $\mu$, prompt fake $\mu$, fake
       $\Jpsi$ and pile-up contributions are represented by dashed
       lines.\label{fig:ThirdmuFit}}
\end{figure}

Having derived templates for each of the expected third muon
components in the data, an extended maximum-likelihood fit is carried
out. The data are fitted in each bin of the various observables with
the signal muon, prompt muon, fake \Jpsi and pile-up templates derived
in the same kinematic bin. Figure~\ref{fig:ThirdmuFit} shows the
result of the third-muon fits to the data for an example bin. While it
can be seen that some of the templates suffer from statistical
fluctuations, these have a very small effect on the fit result, and a
systematic uncertainty is derived to cover these effects, as described
in Section~\ref{sec:thirdmuonunc}.

The stability of the fit is verified with a closure test using \MCd
samples. Pseudo-data sets are created with varying fractions of prompt
and non-prompt third-muon events and the fits repeated. The numbers of
fitted prompt and non-prompt third muon events are compared to the
values used to construct the pseudo-data sets. The differences between
input and fitted values are typically below the 1\% level,
demonstrating that the fits perform well. Two additional qualitative
cross-checks on the modelling of the fake muon component are performed
by considering data control regions with requirements orthogonal to
those of the signal region such that they are expected to contain more
fake muons. The first control region is defined by reversing the
pile-up rejection requirement so that the probability of a charged
pion or kaon faking a third muon increases. The second control region
looking at prompt dimuon events is defined by reversing the
pseudo-proper decay time requirement where the fraction of \Jpsi
production via weak decays is reduced, and the fraction of events
containing a charged pion or kaon faking a third muon is
increased. The fits in these control regions are performed inclusively
as there is insufficient data to split into differential bins.  The
third-muon fit procedure is the same as described above except that no
pile-up template is used in the pile-up control region. In both
control regions, the fit behaves as expected, giving a good
description of the data and returning a higher fraction of fake muons.

\subsection{Irreducible backgrounds}
\label{sec:irreducibles}

There are three additional sources of background that could not be
constrained in the fits, either because their contribution is too
small to be reliably determined, or because their characteristics are
very similar to the signal. As no reliable data-driven determination
is possible, these irreducible backgrounds are instead subtracted from
the post-fit signal yield based on estimates derived from \MCn.

The first source of irreducible background is $B_c\rightarrow
\jpsi+\mu+X$ production. As both the \Jpsi and third muon originate in
the decay of the same hadron, this is considered to be a source of
background, and is concentrated at low values of $\DRy(\jpsi, \mu)$,
the region of particular interest in this analysis. The production
fraction of $B_c$~\cite{ATL-PHYS-PUB-2014-008} and branching
fraction~\cite{Agashe:2014kda} of $B_c\rightarrow\jpsi\mu+X$ mean that
this background is expected to be very small and, in the absence of an
identifiable signal in the data, the estimate is taken directly from
\MCn. A prediction of the $B_c$ contribution passing the event
selection is calculated in each differential observable bin from both
the \pyegt and \hwpp \MCd samples. The average of the two predictions
is then subtracted from the fitted signal yield to remove the
background from $B_c$ decays.

Another source of signal-like muons in this analysis is semileptonic
decays of \Chadrons.  In the case where the \Chadron is produced in
the decay of a \Bhadron, these muons are counted as part of the
signal, but all other \Chadrons are considered as a source of
background muons. This population of events is again expected to be
small, as it requires a displaced \Jpsi produced in a \Bhadron decay,
as well as a separate \Chadron; production modes include separate
$g\rightarrow b\bar{b}$ and $g\rightarrow c\bar{c}$ splittings in the
same hard scatter, or double parton scattering producing $b\bar{b} +
c\bar{c}+X$ in a single proton--proton collision. To determine the
rate of such events in data, it is possible to use the third-muon
\dzerosig fits: \Chadrons have shorter lifetimes than \Bhadrons,
producing a narrower \dzerosig distribution. However, the contribution
is found to be so small that it is not possible to reliably extract
the \Chadron background with the number of events available in each
fit. Instead the rate is derived from simulation, where it is found
that approximately 5\% of all third muons passing the selection
originate from this source of background. To derive a correction for
each bin of the measured kinematic distributions, the rate of
$b$+$c$-hadron production is derived separately from \pyegt and \hwpp
\MCn, two event generators which have different models of the parton
shower, underlying event and double parton scattering, and the average
of these is used to remove the expected contribution from the fitted
number of signal events.

The final background considered is from events where a charged pion or
kaon traverses the detector to the muon spectrometer without
interacting with the detector material or decaying, referred to as
\emph{sail-through}. This background has a signature very similar to
the signal third muons due to the presence of a well-defined track in
the inner detector and a well-matched track in the muon
spectrometer. The estimate of this background is taken from \MCn,
where a reconstructed muon not associated with a muon in the \MC event
record is matched to a charged kaon or pion which has no decay vertex
inside the detector.  The third-muon yield in each differential bin is
corrected, after fitting, by removing the expected number of
sail-through events. This number is estimated from the \pyegt \MCn, by
calculating the ratio of the numbers of sail-through and fake muons
and using this to scale the background estimate from the number of
fake muons fitted in data.

\begin{figure}[h!]
   \centering
        \includegraphics[width=0.7\textwidth]{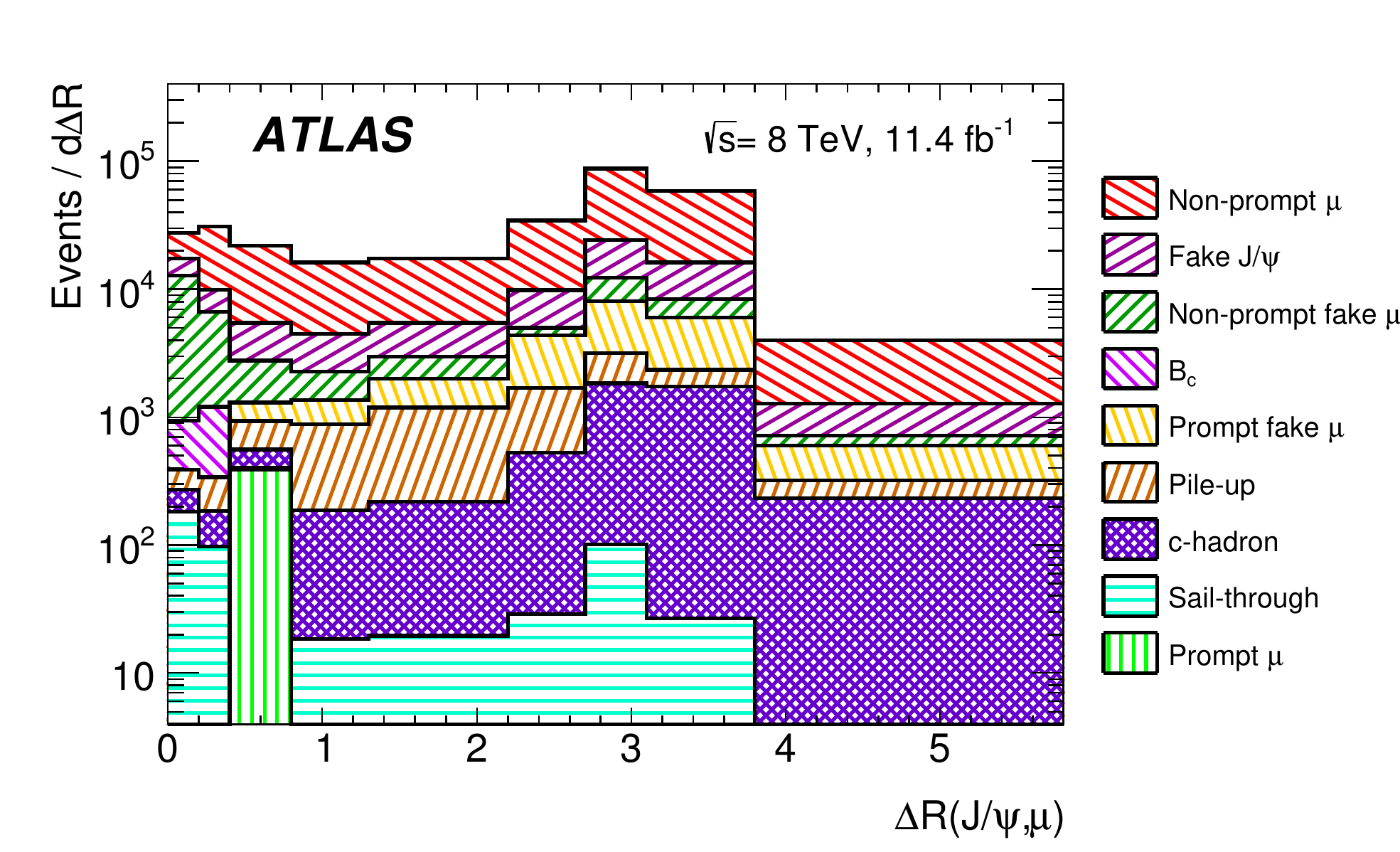}
 \caption{Breakdown of the contributions from the signal and different
   sources of background across the distribution of $\DRy(\Jpsi,
   \mu)$.\label{fig:bkgComp}}
\end{figure}

A summary of the relative sizes and distributions of all the
backgrounds as a function of the $\DRy(\Jpsi, \mu)$ observable is given in
Figure~\ref{fig:bkgComp}. 

\subsection{Extrapolation to the full range of pseudo-proper decay time}
\label{sec:extrapolation}
Once the signal yield has been determined, a correction must be
applied to extrapolate the results obtained in the third muon fit (for
$\tau>0.25$~mm/c) to the full range of \Jpsi pseudo-proper decay time. The
rate of non-prompt third muons is expected to be constant as a
function of $\tau$ for non-prompt \Jpsi mesons, and this is confirmed in
two tests.  First in data, by repeating the third muon fit in
different bins of $\tau$ and observing that the ratio of non-prompt
\Jpsi events to each of the third muon fit components remains
constant. Second with \MCn, where the rate of non-prompt muons is
indeed found to be constant as a function of $\tau$ in events
containing a \Jpsi originating from a \Bhadron decay.

The extrapolation to the full $\tau$-spectrum is then performed by
simply correcting the third muon yield found in the $\tau>0.25$~mm/c
region by an extrapolation factor taken as the ratio of all non-prompt
\Jpsi events over the full $\tau$ range to the number of \Jpsi events
found above the $\tau>0.25$~mm/c requirement, as determined from the
\Jpsi fit. The correction is derived individually for each
differential observable bin, based on the fit in that bin, and is
typically a factor of around 1.9 with no significant kinematic
dependence.

\subsection{Resolution corrections}
\label{sec:unfold}
The final step in converting the measurement to a full particle-level
cross section is to correct for the effects of detector resolution on
the momentum and $\eta$\ of the muons. Detector resolution can have
two main effects, causing events to migrate between bins, or in and
out of the fiducial volume.

Migration between bins can occur when events passing both the
particle-level and detector-level selections are reconstructed in
different bins of the differential cross sections. In this analysis,
the analysis bins are all significantly wider than the detector
resolution, so migrations between bins are a very small effect.

Migrations in and out of the fiducial volume must also be
considered. Detector resolution effects can move an event into or out
of the fiducial region (for example, by migrating individual muons
above or below the muon \pT$>6$~GeV requirement). These effects are
again found to be very small.

The final resolution correction factor combines both the bin-to-bin
and fiducial migration effects, and as both these effects are small, a
simple correction to each bin is sufficient. These corrections are
derived by simply taking the ratio of the particle-level distributions
to the detector-level distributions in the \pyegt sample, where each
sample distribution is derived independently and events are not
required to pass both selections simultaneously. This ratio is then
applied to the data to correct from detector-level quantities to
particle-level quantities. The size of the correction is typically
less than 2\%, although for certain kinematic bins it can be as large
as 5\%.

\section{Systematic uncertainties}
\label{sec:systs}

Various systematic uncertainties are accounted for in this
measurement.  They broadly fit into three categories: uncertainties
associated with the muon efficiency corrections to data, \Jpsi fit
model systematic uncertainties and uncertainties in the background
components in the fits.  Each source of systematic uncertainty is
considered individually by repeating the full differential analysis
from the beginning with the systematic change implemented; the
deviation from the nominal result is then taken as the uncertainty due
to that change. All of the systematic uncertainties, apart from those
concerning \Jpsi modelling, are allowed to vary independently upwards
and downwards; from these two changes, the largest deviation from the
nominal result is symmetrised and assigned as the uncertainty in the
measurement from that source. The total upward or downward systematic
uncertainty in the cross section is then calculated as the sum in
quadrature of all contributions in the upward or downward directions
respectively. When calculating the uncertainty in the normalised
differential cross sections, \emph{shape-only} systematic
uncertainties are derived by varying each source of uncertainty up and
down, while preserving the overall normalisation of the
distribution. The total upwards or downwards systematic uncertainty
for each bin is then the quadrature sum of these shape-only systematic
uncertainties in the upwards or downwards directions respectively.

\begin{figure}[h!]
   \centering
        \includegraphics[width=0.7\textwidth]{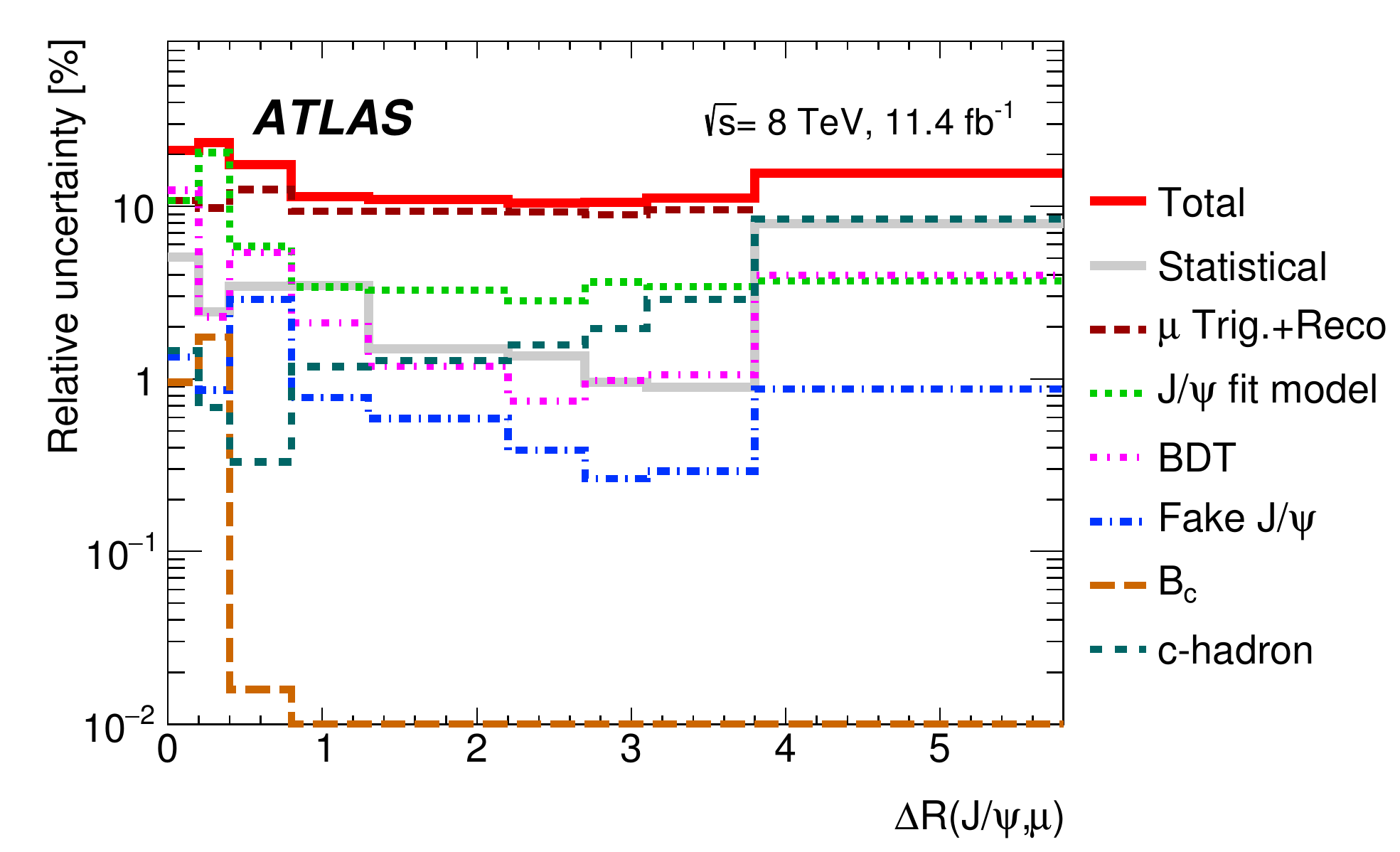}
 \caption{The contribution from each systematic uncertainty source as
   a function of $\DRy(\Jpsi, \mu)$. Details of the different
   contributions are given in the text. \label{fig:JpsiModelSysts}}
\end{figure}

The main sources of systematic uncertainty in the cross section
measurement are shown in Figure~\ref{fig:JpsiModelSysts} for the
$\DRy(\Jpsi,\mu)$ observable.  The statistical uncertainty includes
the statistical uncertainty of the data and the statistical
uncertainty of the third-muon templates taken from \MCn. The trigger
uncertainty includes the uncertainties in the single-muon trigger
efficiency, the dimuon efficiency correction, the nearby-muon
correction and the \MCn-based correction to the trigger
efficiency. The muon reconstruction uncertainty includes the
uncertainties in the muon reconstruction efficiency and the inner
detector track reconstruction efficiency. The fit model uncertainty
includes several variations of the functional forms used in the fit
model and the fitting procedure. The BDT uncertainty includes several
uncertainties on the \MCn-derived templates and a data-driven
uncertainty. The Fake~\Jpsi, $B_c$ and $c$-hadron uncertainties are
included as described in Section~\ref{sec:thirdmuonunc}. The
uncertainties in the resolution correction, pile-up and pile-up double
counting, also described in Section~\ref{sec:thirdmuonunc}, are
omitted from the figure for clarity but are included in all following
calculations. The luminosity uncertainty is constant and therefore
also not included in the figure.

All these uncertainties are described in more detail in the remainder of
this section.

\subsection{Luminosity uncertainty}

A 1.9\% uncertainty is assigned to the delivered integrated
luminosity. The methodology used to determine this uncertainty is
described in Ref.~\cite{Aaboud:2016hhf}.

\subsection{Muon trigger and reconstruction efficiency uncertainties}

As described in Section~\ref{sec:muontrigreco}, the trigger efficiency
has several components: the single-muon efficiency; the efficiency to
identify pairs of muons with opposite-charge and vertex requirements;
and the efficiency to resolve pairs of nearby muons including a
third-muon correction.

The single-muon efficiency is parameterised as a function of \pt and
$q \cdot \eta$. Each bin in the trigger efficiency map has an
associated uncertainty resulting primarily from the limited data
available to derive the efficiency. A Gaussian p.d.f. is formed for
each map bin, with the mean being given by the central value and the
width by the uncertainty in that bin. Modified efficiency maps are
formed by sampling randomly from the p.d.f. in each map bin; multiple
maps can be created by repeating this procedure. The data are
corrected for trigger efficiency using each modified map in turn to
determine the number of events in the reweighted data set after
applying the trigger efficiency correction.  The distribution of the
corrected data yields using all the maps is then fitted with a
Gaussian function; the mean of this Gaussian function gives the
nominal event yield after corrections, and the width gives the
systematic uncertainty in the trigger efficiency weighting procedure.

The trigger efficiency map is also corrected by a factor derived from
\MCn to remove the bias in the maps due to the tag-and-probe threshold
being different to that of the analysis selection, and a systematic
uncertainty in this correction is defined by using the data-driven
maps without the \MCn-based correction applied and repeating the
analysis. The change in the extracted differential cross sections from
using the uncorrected maps is taken as a systematic uncertainty.

The efficiency to identify pairs of muons with opposite-charge and
vertex requirements also has an uncertainty, and the impact of this is
assessed separately by varying the nominal correction for this
efficiency by one standard deviation when reweighting the data set. In
addition, an uncertainty assigned to the efficiency correction for
cases where the third muon is close to a trigger muon is assigned by
varying the correction parameters within their uncertainties.

The uncertainty in the muon reconstruction efficiency is factorised
into two components. A constant 0.5\% uncertainty is included for the
efficiency of reconstructing a muon track in the inner
detector~\cite{Aad:2014rra}. This is added coherently for each of the
three muons in an event, resulting in a 1.5\% systematic
uncertainty. As in the case of the trigger map efficiency, the
uncertainty in the muon reconstruction maps is defined by the spread
of the data set yields when using a set of modified maps created by
sampling from p.d.f.s defined in each bin of the efficiency map by the
nominal efficiency and the uncertainty in that value.

The combined systematic uncertainty in the reconstruction and trigger
efficiencies on the cross section is 10\%. The relative normalised
uncertainty can be as large as 10\% in certain kinematic regions. This
is dominated by the uncertainty in the dimuon correction to the single
trigger efficiency except in the small $\Delta\phi(\Jpsi,\mu)$ and
$\DRy(\Jpsi,\mu)$ regions where the systematic uncertainty due to a
third muon being close to a triggered muon is largest.

\subsection{\Jpsi mass--lifetime model uncertainty}

To assess potential bias in the fitted number of non-prompt \Jpsi
candidates extracted from the two-dimensional dimuon fit due to the
models used, various changes are made to the functions describing the
fit components:
\begin{itemize}
\item The \Jpsi mass model is changed to a
combination of two Gaussian functions.  
\item The non-prompt \Jpsi pseudo-proper decay time model is changed
  to a double exponential function convolved with the same resolution
  function.
\item The resolution model is changed to a single Gaussian function.
\item The fixed parameters in the Crystal Ball function are varied by
  $\pm10\%$.
\item The dimuon mass model of the prompt and double-sided fake \Jpsi
  backgrounds is changed to be a first-order polynomial function.
\item The single-sided fake \Jpsi background dimuon mass model is
  changed to a first order polynomial function.
\item The single-sided fake \Jpsi pseudo-proper decay time model is
  changed to a double exponential function.
\end{itemize}

The analysis is repeated for each of the varied \Jpsi models, with
only one change at a time. When performing fits to lower statistics
regions, the fits are naturally less constrained, leading to a
potential double counting of statistical uncertainties when performing
the varied fits. To avoid this, the envelope of the largest deviation
from the nominal event yield when considering all the individual model
changes is taken as the total systematic uncertainty for the \Jpsi
model uncertainty.  This envelope is calculated separately in each
differential bin.

The fit model variations give a 5\% uncertainty in the total cross
section and at most a 15\% relative uncertainty in the normalised
differential cross section. This is dominated by the change in decay
time parameterisation of the non-prompt \Jpsi mesons and the mass
parameterisation of the single-sided fake \Jpsi background.

\subsection{Third-muon uncertainties\label{sec:thirdmuonunc}}

There are several contributions to the systematic uncertainty in the
third-muon fit from the derivation of the templates and the estimation
of the various background components.

There is a statistical uncertainty in the templates used for the
third-muon fits, and the effect of this uncertainty is assessed
through an ensemble test.  Pseudo-templates are created by randomly
sampling from the default templates, and the fit repeated using each
of the modified pseudo-templates. This procedure provides a
distribution of results for the extracted signal muon yield, and this
distribution is fitted with a Gaussian function, the width of which is
taken as a systematic uncertainty from the statistical fluctuations of
the templates. The uncertainty is typically at the level of
1--2\%. The shapes of the templates derived from \MCn are found to
have minimal dependence on the physics modelling in the \MCn, due to
the similar lifetimes of all \Bhadrons and the very weak kinematic
dependence of \dzerosig, so no additional systematic uncertainty was
required.

The contribution from events with a fake \Jpsi candidate which is
combined with a third muon from another proton--proton interaction is
double-counted by the fitting process. It would be possible to remove
this double-counting by creating a four-dimensional fit but the
complexity and lack of stability in such a fit make this very
impractical and, since the level of double counting is very small, this
is not necessary. Instead, each kinematic bin can be corrected by
determining the number of fake \Jpsi events that are due to pile-up
and removing that from the number of fake \Jpsi events extracted from
the fit. The number of fake \Jpsi events that are due to pile-up is
determined by fitting the $\Delta z_{0}$ distribution of events
outside the signal mass window. Approximately 2\% of fake \Jpsi events
are found to be from pile-up events and the fake \Jpsi contribution is
of order 10\% of the selected data, therefore the signal yield is
typically altered by 0.2\%, but varies depending on the kinematic region.

The fake muon template contains two types of background with similar
behaviour: DIF and hadronic leakage, as explained in
Section~\ref{sec:fitThirdMu}. Both backgrounds are due to the decays
or interactions of charged pions and kaons. To assess the robustness
of the \MCn of the fake muon background, the templates used in the
third muon fits are systematically altered.  The BDT response is
subtly different for pions and kaons but, due to the limited number of
fake muon candidates in \MCn, the two sources of fake muons are
combined. The ratio of pions to kaons populating the fake muon
templates is changed by $\pm50\%$ to cover any effect of the
combination.  In addition, the fraction of pions and kaons decaying
inside and outside the inner detector is varied by $\pm50\%$.  Finally
the ratio of DIF muons and hadronic leakage muons in the fake muon
template is changed by $\pm50\%$.  The BDT response is different for
the two types of fake muons but the available number of \MCd events do
not allow separation of the two contributions in the template which is
composed of approximately 75\% DIF muons. The fractional composition
of the fake muon template is changed to cover any mismodelling in \MCn
of the composition of the two sources of fake muons. The effects these
systematic shifts have on the BDT template result in an uncertainty of
less than 1\% in the total cross section and up to 2\% relative
uncertainty in the differential normalised cross section in certain
kinematic regions.

Furthermore, a data-driven uncertainty in the shape of the BDT
distribution for fake muons is derived using two control regions.  A
pile-up region is defined by reversing the $|\Delta z_{0}|$
requirements, and a fake \Jpsi region is formed by selecting for
events outside the dimuon mass window.  The third-muon BDT
distribution in these events is fitted with the usual \MCn-derived BDT
templates for real and fake muons. The fitted real component is then
subtracted from data, leaving an estimate of the BDT template for fake
muons.  Due to statistical limitations, this process can only be
performed on the inclusive data set and not differentially.  This
data-driven template is then used instead of the \MCn-derived fake
muon templates for each differential fit.  The difference with respect
to the usual result is then used to form the systematic uncertainty.
The envelope from the largest deviation in each bin from either the
pile-up or fake \Jpsi derived template is assigned as a systematic
uncertainty in the modelling of fake muons in \MCn. The corresponding
uncertainty in the total cross section is 1\% and can be as large as a
10\% relative uncertainty in the normalised differential cross
sections.

The uncertainty in the fake \Jpsi background estimate is assessed by
changing the normalisation of the fake \Jpsi templates in the
third-muon fits.  The number of fake \Jpsi events is derived from the
two-dimensional dimuon fits, given by normalisation of the three fake
\Jpsi components within the dimuon signal mass window.  Due to the
dimuon pseudo-proper decay time requirement in the third-muon fit region, the
single-sided fake \Jpsi component is the only background to contribute
non-negligibly to the high pseudo-proper decay time region.  The
differential fits are repeated with fake \Jpsi template normalisation
altered by the uncertainty ($\pm1\sigma$) in the normalisation of the
single-sided background. The systematic uncertainty assigned is less
than 1\% on the total cross section but as much as 3\% relative
uncertainty in the normalised differential cross sections.

For the uncertainty in the pile-up background, a similar procedure is
used.  The templates used in the third-muon fits are changed by
altering the normalisation within the uncertainty.  The uncertainty is
derived from the Gaussian fit to $\Delta z_{0}$ and is applied as a
$\pm1\sigma$ variation to the nominal pile-up templates. The
corresponding systematic uncertainties are very small, 0.4\% in the
total cross section and up to 2\% in the normalised differential cross
sections.

The $B_{c}$ background prediction is taken from the average of \pyegt
and \hwpp \MCn predictions.  The difference between the two
predictions is assigned as an uncertainty in the number of
$B_c$-hadrons in the data set. Similarly the number of events
estimated to be from $b$+$c$-hadrons is taken from the average of
\pyegt and \hwpp \MCn predictions.  A systematic uncertainty is
assigned to this prediction, using the difference between \pyegt and
\hwpp for the rate of $b$+$c$-events. The systematic uncertainty
assigned to the $B_c$-hadron prediction gives a small uncertainty of
below 0.1\% on the total cross section. Typically the relative
uncertainty in the normalised differential cross section is also very
small, except at low $\DRy(\Jpsi, \mu)$ where it can be almost
2\%. For the $b$+$c$-hadron estimation, the systematic uncertainty in
the total cross section is 2\% and the relative uncertainty in the
normalised differential cross section can be as large as 9\%.

The sail-through background prediction is difficult to constrain with
data, so the \MCn-based estimate is varied by $\pm50\%$ to assign an
uncertainty due to mismodelling. The corresponding uncertainty in the
total cross section is 0.1\% and the relative uncertainty in the
normalised differential cross section is less than 1\% everywhere.

\subsection{Resolution correction uncertainty}

The uncertainty in the factors used to correct for events migrating in
and out of the acceptance is estimated based on the statistical
uncertainty in the \MCd sample used to derive the correction. The
correction, described in Section~\ref{sec:unfold}, is derived from the
ratio of events passing the particle-level selection to events passing
the detector-level selection in \MCn.  The uncertainty in this ratio
is calculated assuming these samples are uncorrelated. That is not
entirely the case as they are derived from the same \MCd sample, so
this represents a conservative estimate of an uncertainty in this
correction. The relative fractional uncertainty due to the resolution
corrections for $\Delta\phi(\Jpsi, \mu)$ is typically at the 1\%
level.

\section{Results and interpretation}
\label{sec:result}

The total measured cross section in the fiducial region, defined in
Section~\ref{sec:fiducial}, is
\begin{equation*}
\sigma(\B(\rightarrow\Jpsi[\rightarrow\mumu]+X)\B(\rightarrow\mu+X)) =
17.7 \pm 0.1 (\textrm{stat}) \pm 2.0 (\textrm{syst})~\textrm{nb.}
\end{equation*}
While leading order calculations are not expected to accurately reproduce
this total cross section, the normalised differential cross sections
are used to test the accuracy of a number of predictions. First,
comparisons are made using \pyegt, exploring several different options
for the $g\rightarrow b\bar{b}$ splitting kernel, as this process
dominates the region of particular interest: small-angle \Bhadron
production.  The details of these settings are given in
Ref.~\cite{PY8gbb} and summarised in Table~\ref{tab:py8opts}. The
settings explore one of the main theoretical degrees of freedom when
evaluating gluon-splitting to heavy quarks: whether to use the
relative \pt (Opt. 1 and 4) or mass (Opt. 5, 8, 5b and 8b) of the
splitting to set the scale when determining the value of
$\alpha_{\textrm{s}}$\ to be used in that splitting.

\begin{table}[h!]
\begin{center}
\begin{tabular}{ p{1.3cm} p{14cm} }
\hline Option label & Descriptions \\ \hline Opt. 1 & The same
splitting kernel, $(1/2) (z^2 + (1-z)^2)$, for massive as massless
quarks, only with an extra $\beta$ phase-space factor. This was the
default setting in \pyegt.1, and currently must also be used with the
\mcatnlo~\cite{1126-6708-2002-06-029} method.\\ Opt. 4 & A splitting
kernel $z^2 + (1-z)^2 + 8r_q z(1-z)$, normalised so that the
$z$-integrated rate is $(\beta/3) (1 + r/2)$, and with an additional
suppression factor $(1 - m{_{qq}}^2/m_{\textrm{dipole}}^2)^3$, which
reduces the rate of high-mass $q\bar{q}$~pairs. This is the default
setting in \pyegt.2.\\ Opt. 5 & Same as Option 1, but reweighted to an
$\alpha_{\textrm{s}}(k m_{qq}^2)$ rather than the normal
$\alpha_{\textrm{s}}(\pt^2)$, with $k=1$.\\ Opt. 5b & Same as Option
5, but setting $k=0.25$.\\ Opt. 8 & Same as Option 4, but reweighted
to an $\alpha_{\textrm{s}}(k m_{qq}^2)$ rather than the normal
$\alpha_{\textrm{s}}(\pt^2)$, with $k=1$. \\ Opt. 8b & Same as Option
8, but setting $k=0.25$.\\ \hline
\end{tabular}
\caption{Description of \pyegt options. Options 2, 3, 6 and 7 are less
  well physically motivated and not considered here. The notation used
  is as follows: $r_q = m_q^2/m_{qq}^2$, $\beta = \sqrt{1 - 4r_q}$,
  with $m_q$ the quark mass and $m_{qq}$ the $q\bar{q}$ pair invariant
  mass. } \label{tab:py8opts}
\end{center}
\end{table}

\begin{figure}[htbp]
  \centering 
  \includegraphics[width=0.49\textwidth]{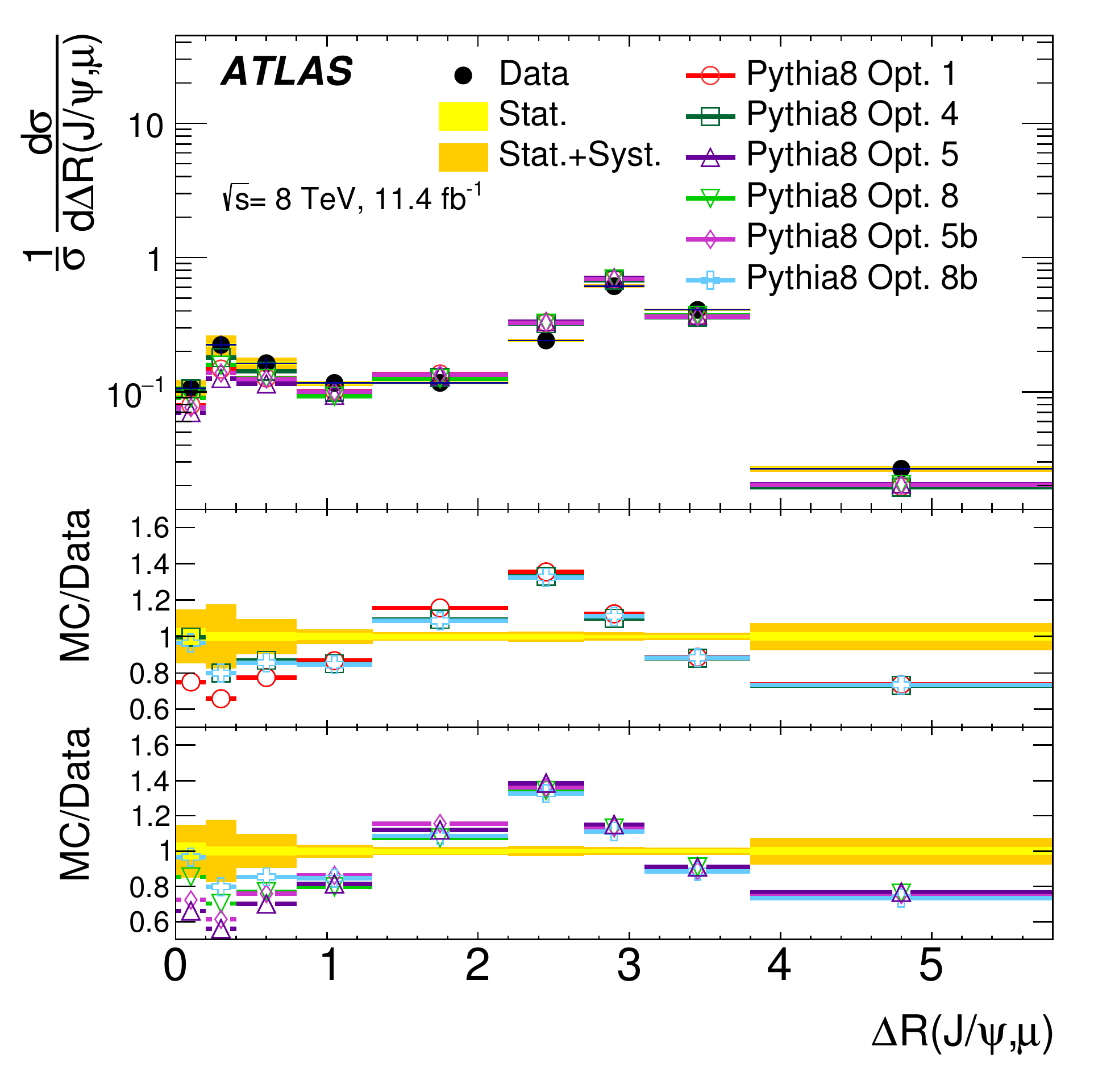} 
  \includegraphics[width=0.49\textwidth]{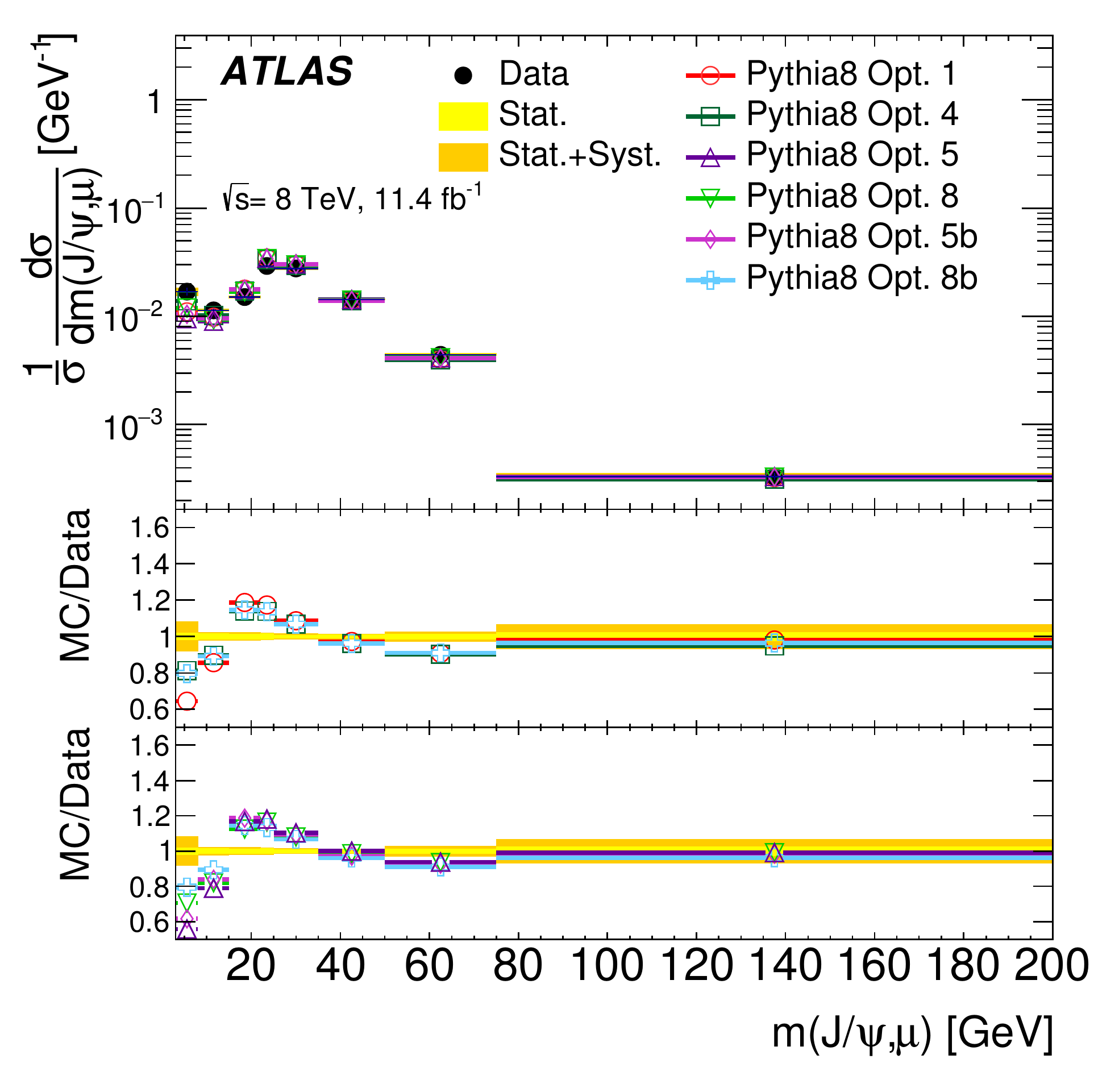}
  \includegraphics[width=0.49\textwidth]{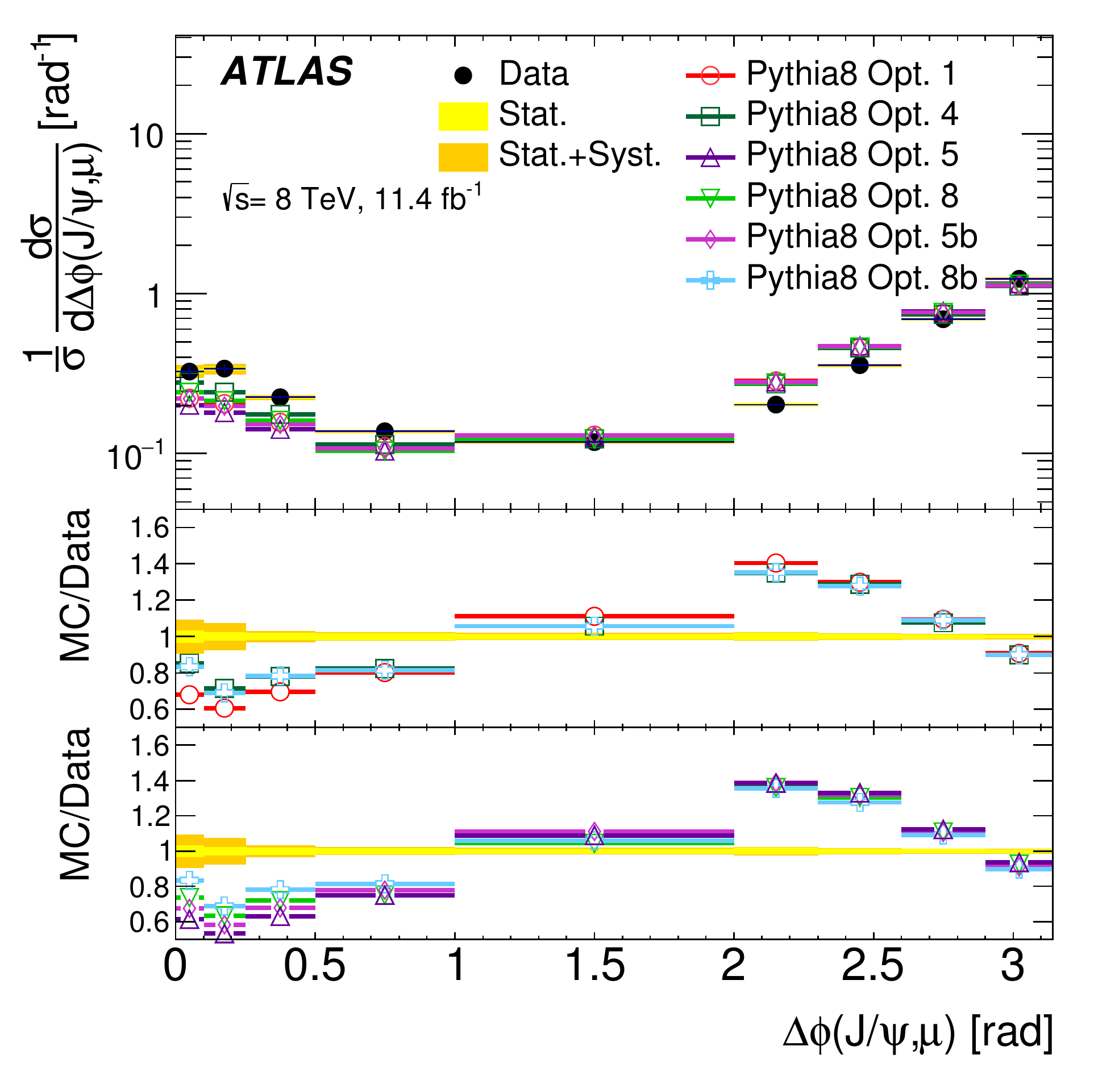}
  \includegraphics[width=0.49\textwidth]{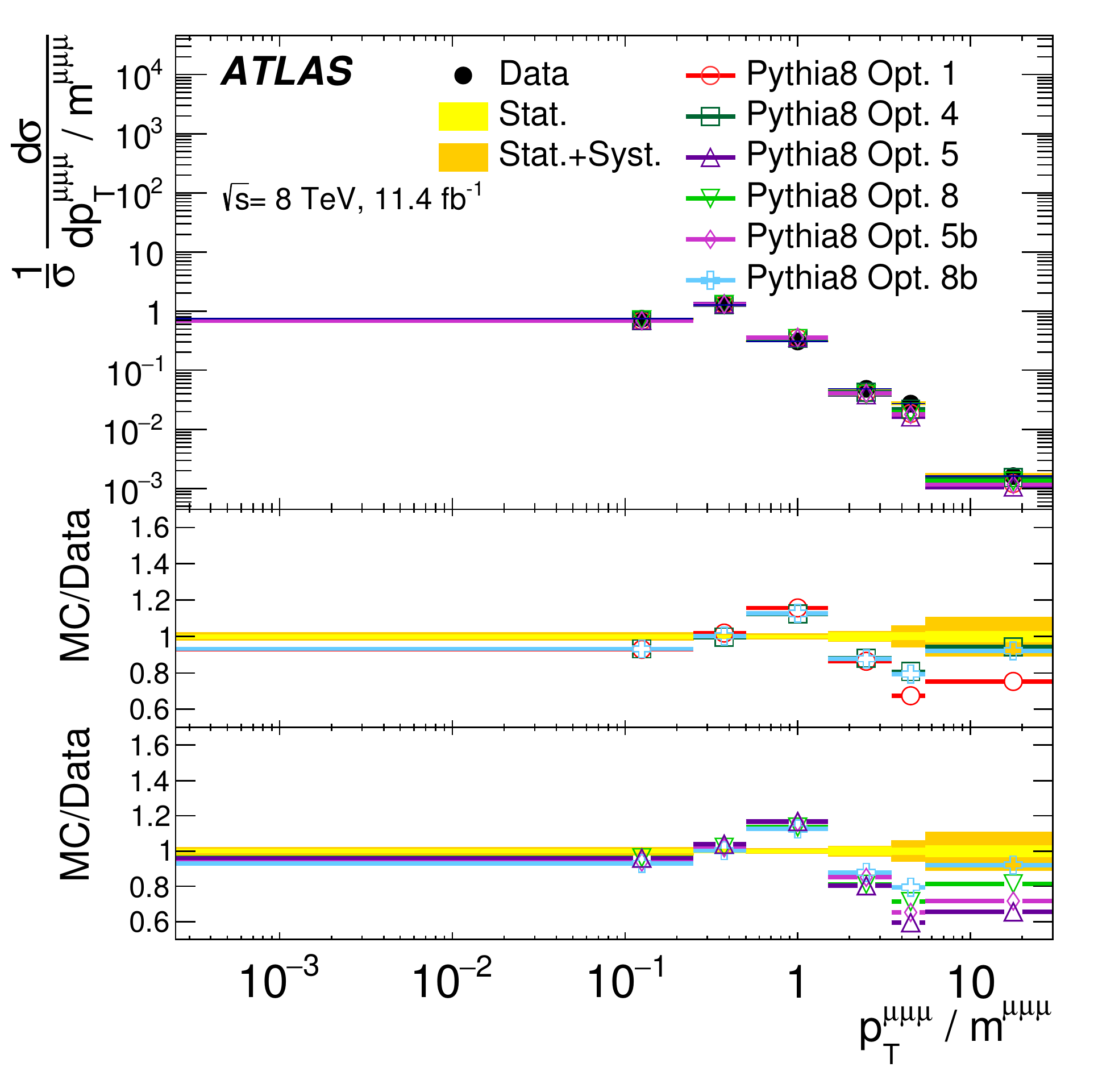}
  \caption{Measured normalised differential cross sections as a
    function of $\DRy(\Jpsi, \mu)$, $m(\Jpsi, \mu)$, $\Delta
    \phi(\Jpsi, \mu)$ and $\pT/m$ compared to \pyegt predictions with
    the different gluon-splitting parameter settings described in
    Table~\ref{tab:py8opts}. The bottom pane shows the ratio to data
    of the options that use a mass-based splitting scale. The middle
    pane shows the \pt-based scale options, along with the mass-based
    option that agrees best with data (Opt. 8b) for comparison. }
  \label{fig:JpsiMuResults2}
\end{figure}

The distributions of $\DRy(\Jpsi, \mu)$, $m(\Jpsi, \mu)$, $\Delta
\phi(\Jpsi, \mu)$ and $\pT/m$ are shown in
Figure~\ref{fig:JpsiMuResults2}. In general, \pyegt does not reproduce
the shape of the angular distributions in data within
uncertainties. The \pt-based scale splitting kernels (Opt. 1 and 4)
generally give a better description of the low $\DRy(\Jpsi, \mu)$
region, with the kernel of Opt. 4 performing the best. This region is
more suppressed in the mass-based scale kernels, although this
suppression is overcome when lowering the scale by a factor of four
(Opt. 5b and 8b), with Opt. 8b in particular performing comparably to
Opt. 4. This pattern is repeated across the other differential cross
sections considered.

To extend the comparisons, the \hwpp sample described in
Section~\ref{sec:dataset} is included.  Two samples are also simulated
using \mgamc{}v2.2.2~\cite{Alwall:2014hca} at leading order in QCD
interfaced to the \pyegt.186 parton shower model. In both samples, the
\textsc{CKKW-L}~\cite{Lonnblad:2001iq,Lonnblad:2011xx} merging
procedure is applied, with a merging scale of 15~GeV. The
A14~\cite{ATL-PHYS-PUB-2014-021} tuned parameter settings are used
together with the \textsc{NNPDF2.3LO} PDF set~\cite{Ball:2012cx} for
\pyegt. The \evtgen{}1.2.0 program~\cite{EvtGen} is used for
properties of the \Bhadron and \Chadron decays.  Both samples are
generated with up to one additional parton in the matrix element
calculation. One is generated in the \emph{5-flavour scheme} where
massless $b$-quarks are included in PDF (the
\textsc{NNPDF3.0NLO}~\cite{Ball:2014uwa} PDF set is used) and as
possible initial state partons in the matrix element calculation; this
is referred to as \emph{5fl} in the figures. The other is generated in
the \emph{4-flavour scheme} where $b$-quark mass is included in the
calculation, but $b$-quarks are excluded from the PDF (the
\textsc{NNPDF3.0NLO 4fl} PDF set is used), but are generated in the
matrix element; this is referred to as \emph{4fl} in the figures.

Finally, a sample is simulated using the
\sherpa{}2.1.1~\cite{Gleisberg:2008ta} event generator. Matrix
elements are calculated with two or three outgoing partons at leading
order, and merged with the \sherpa parton
shower~\cite{Schumann:2007mg} using the \textsc{ME+PS@LO}
prescription~\cite{Hoeche:2009rj}. The
\textsc{CT10nlo}~\cite{Lai:2010vv} PDF set is used in conjunction with
dedicated parton shower tuning developed by the \sherpa authors. Only
the 5-flavour scheme sample is considered for the \sherpa event
generator.

Due to the computational demands in producing sufficient events in the
three-muon fiducial volume, both the \mgamc{}+{}\pyegt and \sherpa
samples were produced only to the level of two \Bhadrons. These
predictions are corrected to a three-muon prediction using the
\emph{transfer functions} described in Appendix~\ref{sec:transfer};
the theoretical uncertainty in these transfer functions completely
dominates the statistical uncertainty of the samples, and the
quadrature sum of these two uncertainties is shown as an uncertainty
bar for the \mgamc{}+{}\pyegt and \sherpa samples in the figures.

\begin{figure}[htbp]
  \centering 
  \includegraphics[width=0.49\textwidth]{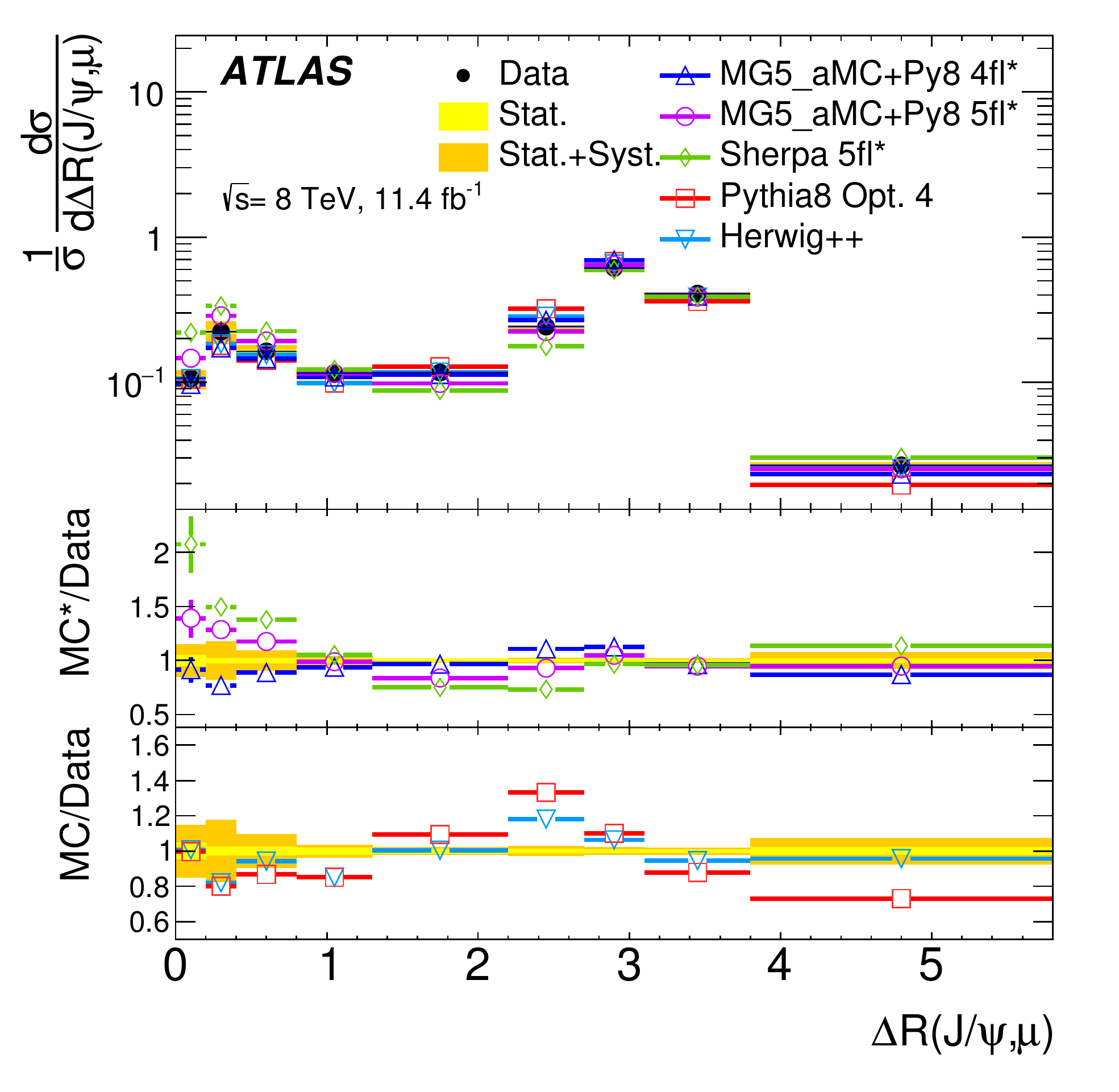} 
  \includegraphics[width=0.49\textwidth]{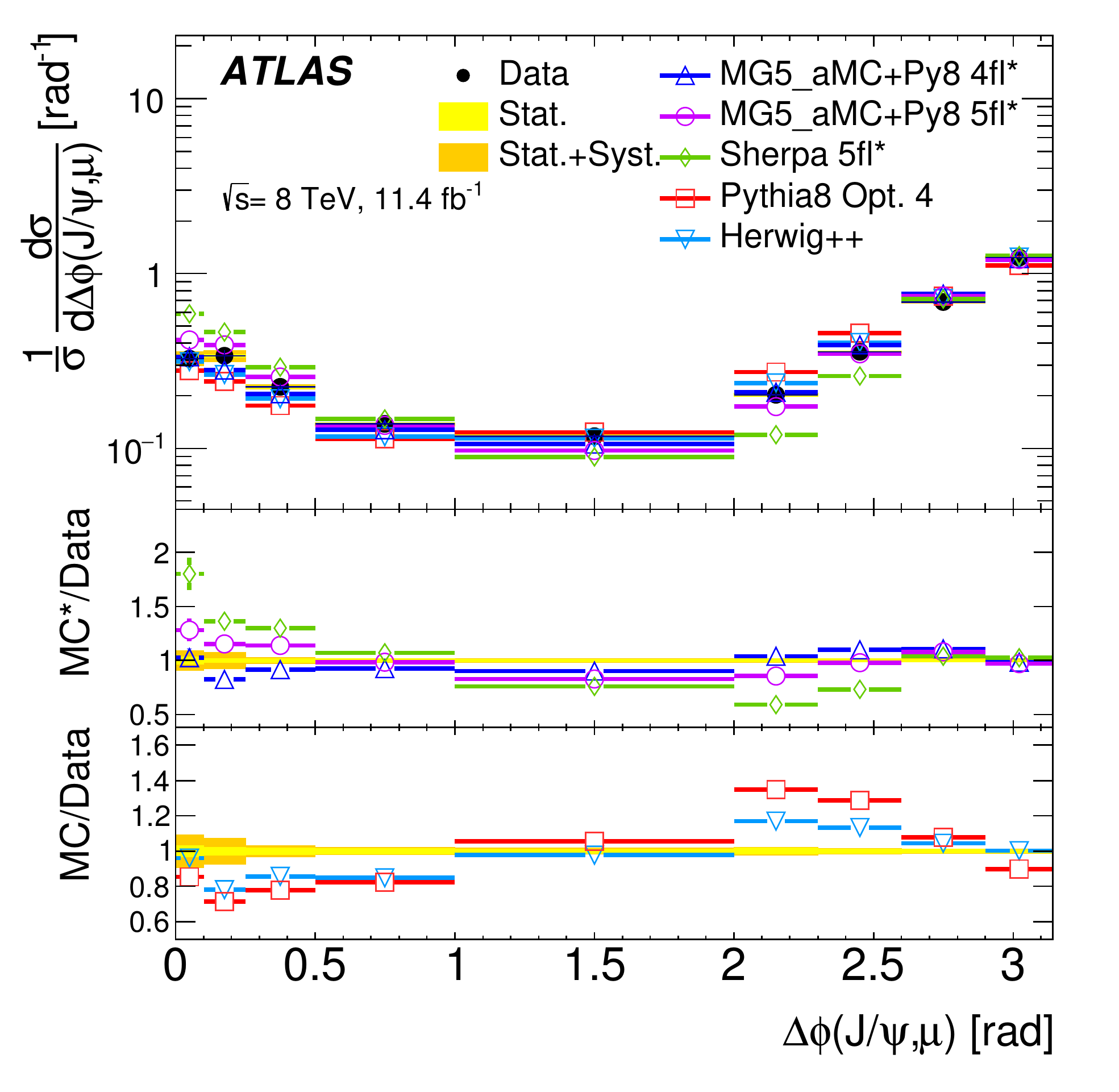}
  \includegraphics[width=0.49\textwidth]{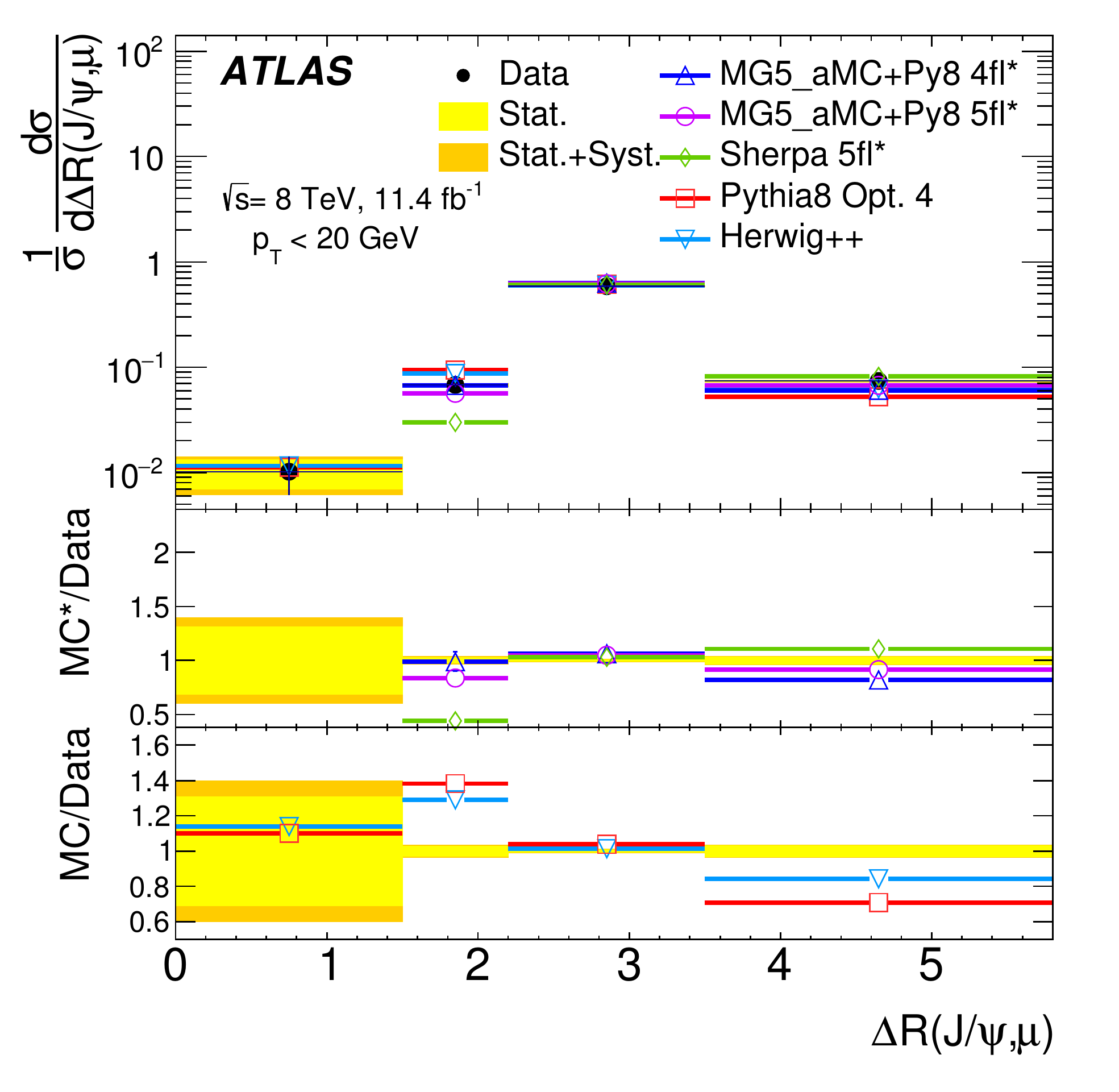} 
  \includegraphics[width=0.49\textwidth]{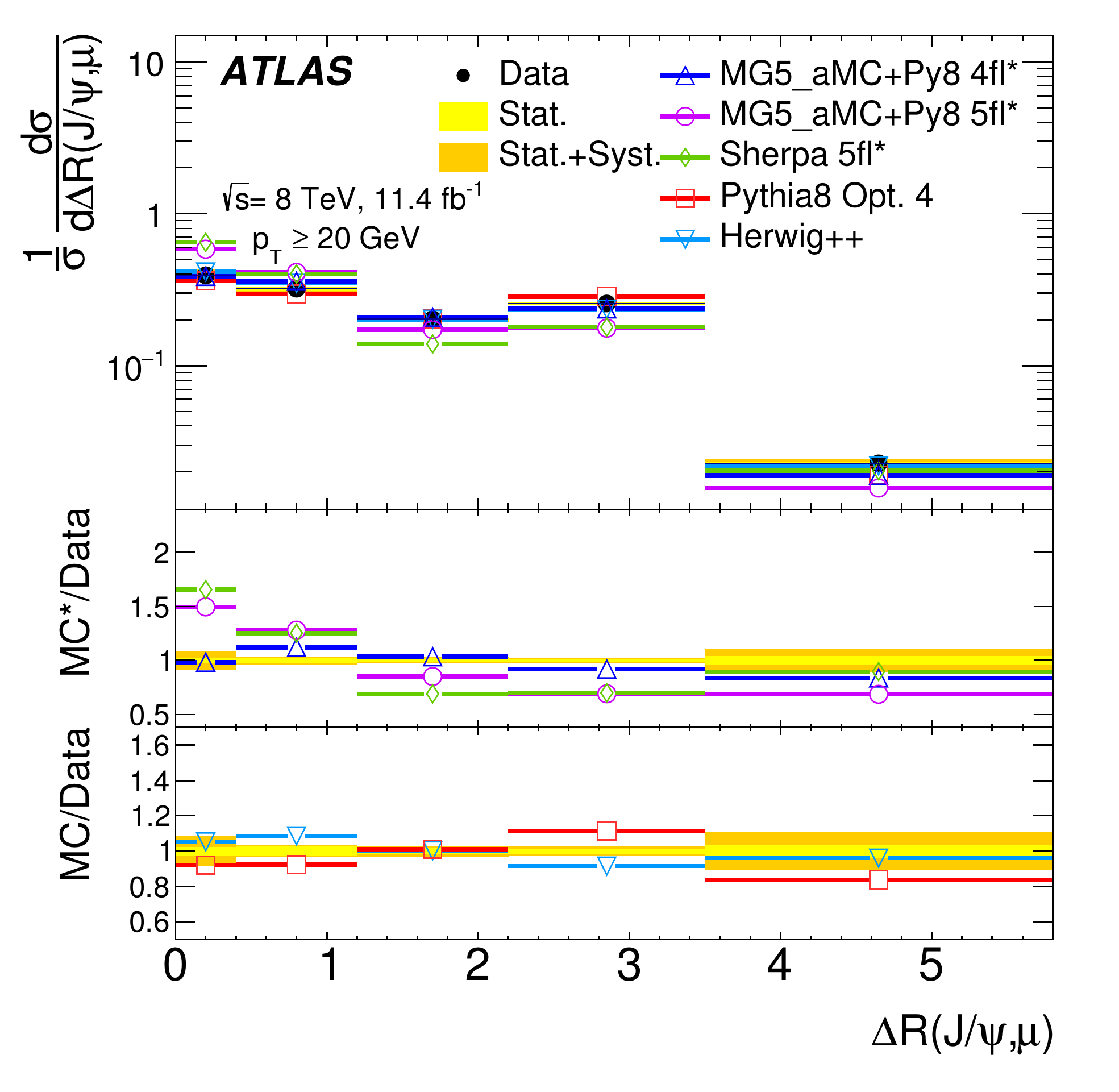}

  \caption{Measured normalised differential cross sections as a
    function of $\DRy(\Jpsi, \mu)$, $\Delta \phi(\Jpsi, \mu)$, low-
    and high-$\pT$ $\DRy(\Jpsi, \mu)$. Comparisons are made with
    predictions of \pyegt and \hwpp. \mgamc{}+{}\pyegt and \sherpa
    predictions are also compared having been corrected from the
    two-\Bhadron production to the three-muon final state via transfer
    functions (indicated with *). There is no entry for these
    predictions in the lowest low-$\pT$ $\DRy(\Jpsi, \mu)$ bin as the
    transfer function is not defined in this bin. The \pyegt
    ``Opt. 4'' gluon-splitting parameter settings are described in
    Table~\ref{tab:py8opts}. The ratio to data of the
    \mgamc{}+{}\pyegt and \sherpa (middle pane), and \pyegt and \hwpp
    (bottom pane) are also shown.}
  \label{fig:JpsiMuResults3}
\end{figure}

\begin{figure}[htbp]
  \centering 
  \includegraphics[width=0.49\textwidth]{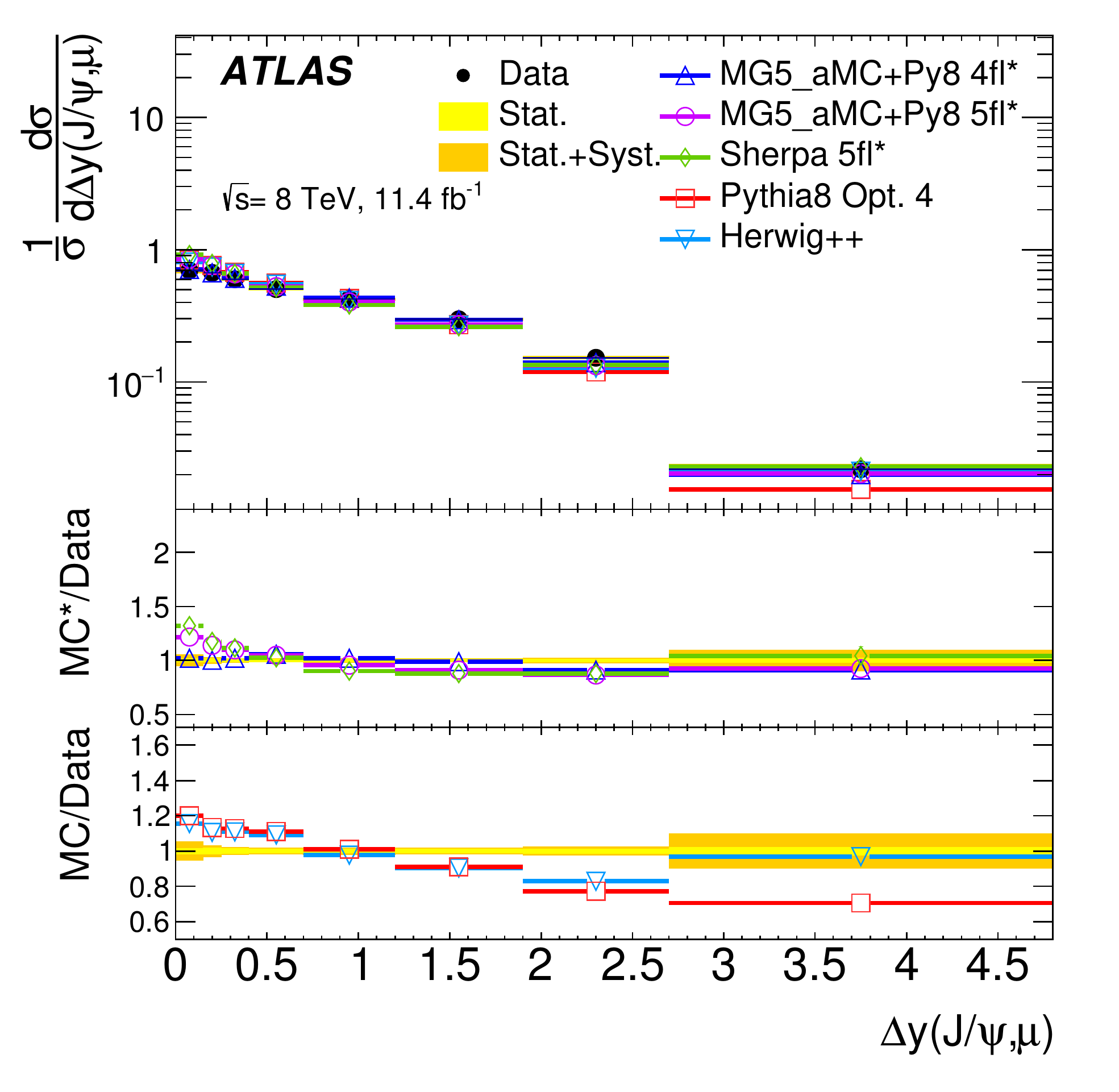}
  \includegraphics[width=0.49\textwidth]{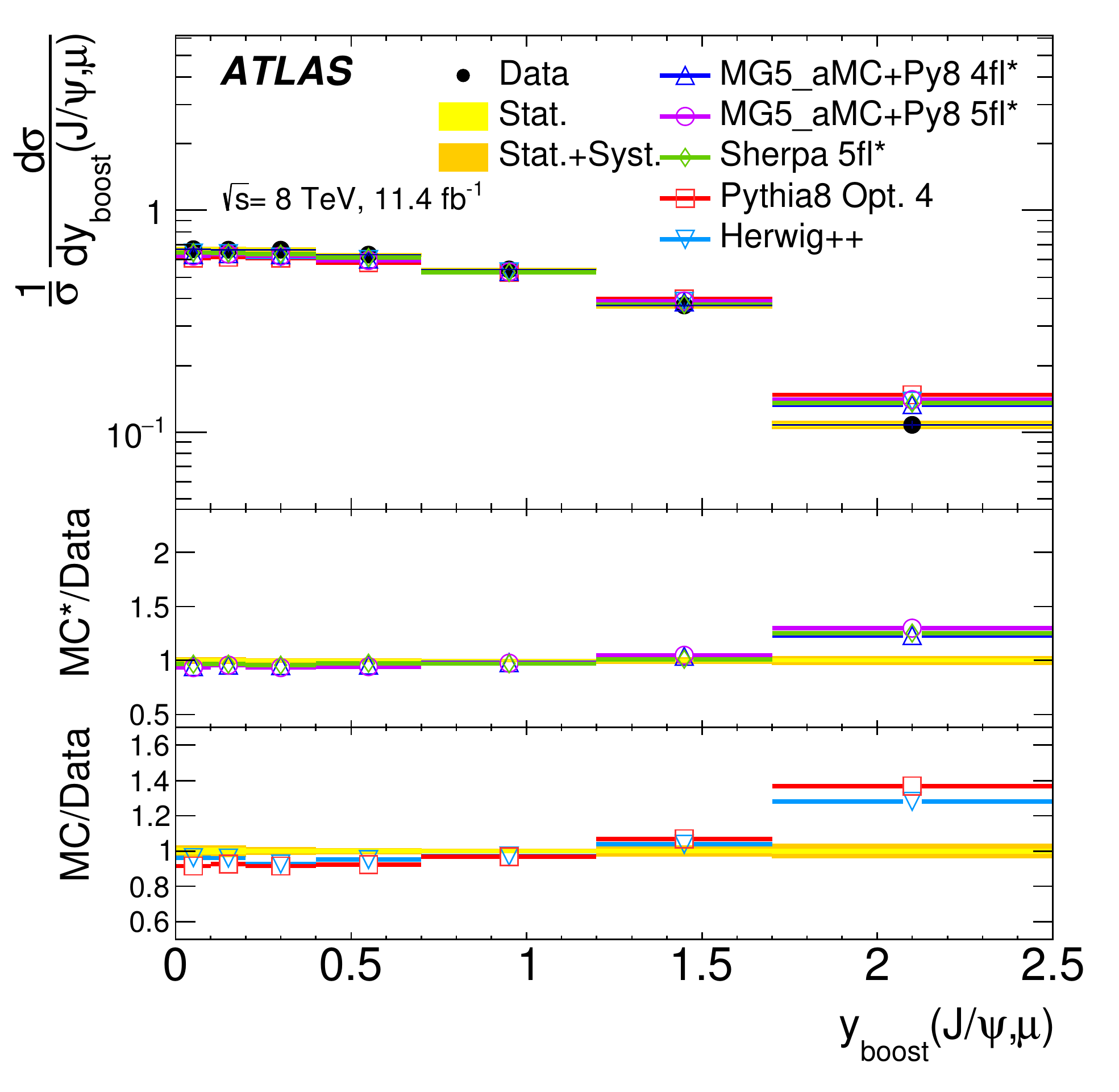}
  \includegraphics[width=0.49\textwidth]{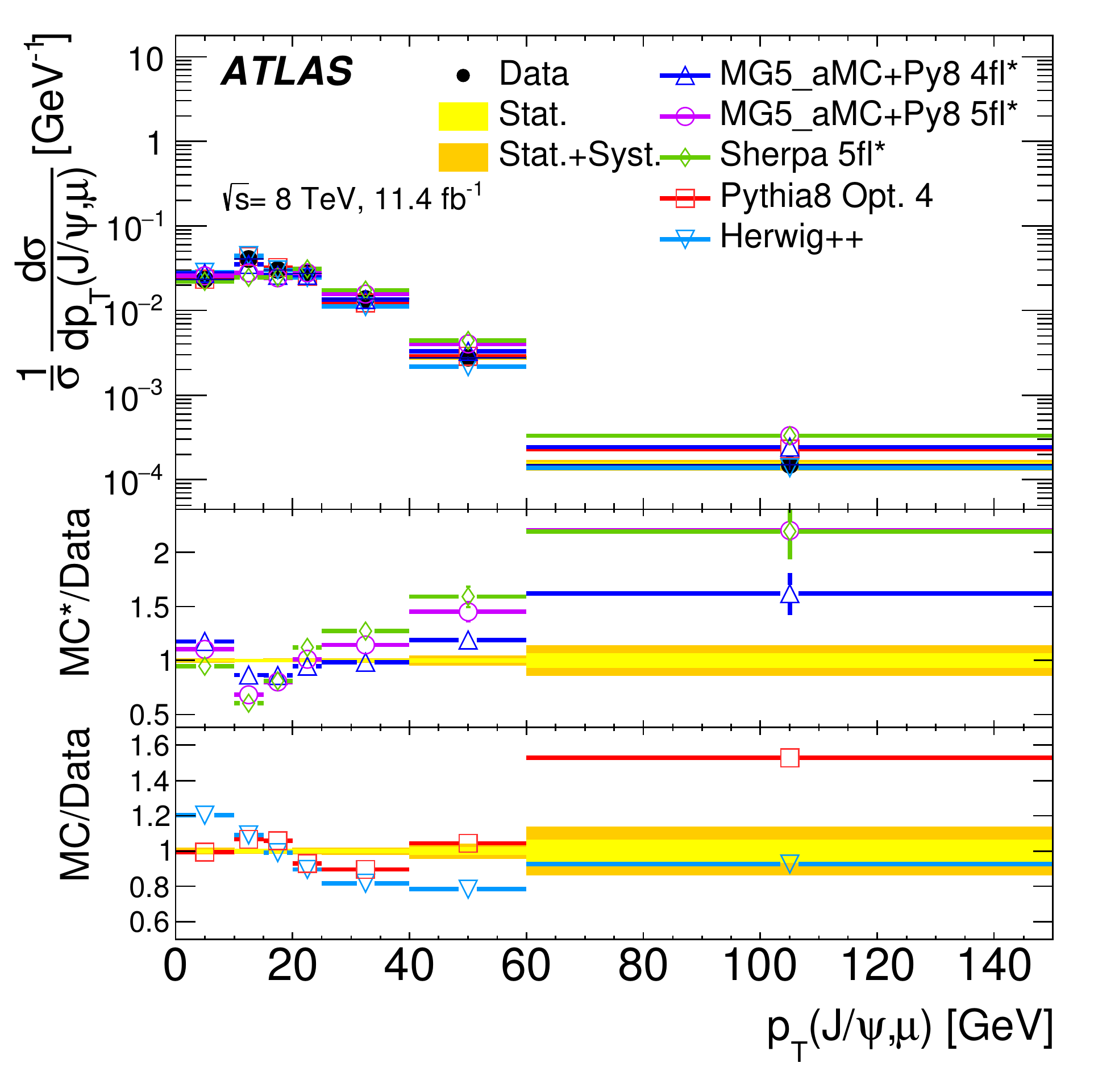}
  \includegraphics[width=0.49\textwidth]{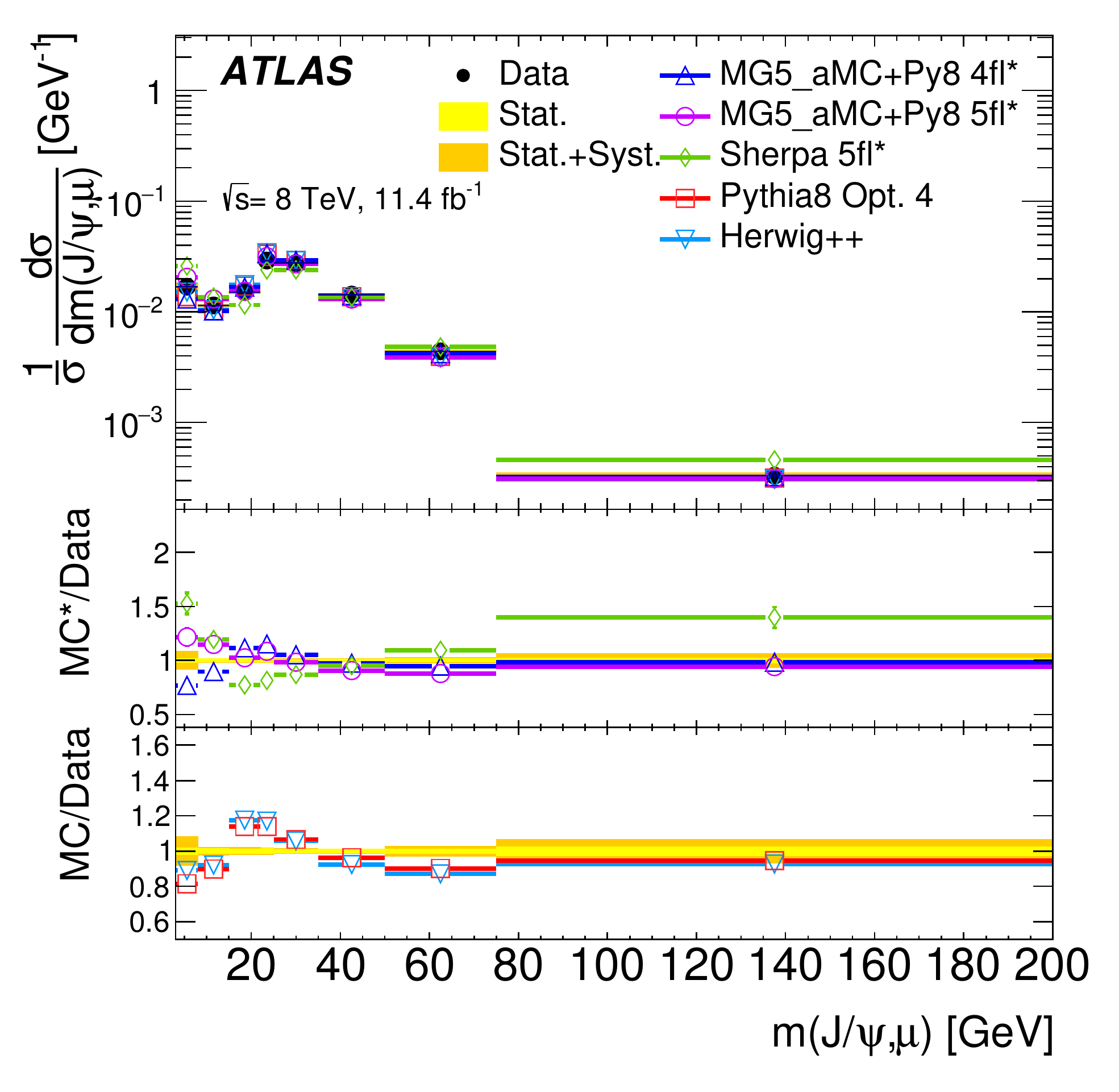}
  \caption{Measured normalised differential cross sections as a
    function of $\Delta y(\Jpsi, \mu)$, $y_{\textrm{boost}}$,
    $\pT(\Jpsi, \mu)$ and $m(\Jpsi, \mu)$. Comparisons are made with
    predictions of \pyegt and \hwpp. \mgamc{}+{}\pyegt and \sherpa
    predictions are also compared having been corrected from the
    two-\Bhadron production to the three-muon final state via transfer
    functions (indicated with *). The \pyegt ``Opt. 4'' gluon
    splitting parameter settings are described in
    Table~\ref{tab:py8opts}. The ratio to data of the
    \mgamc{}+{}\pyegt and \sherpa (middle pane), and \pyegt and \hwpp
    (bottom pane) are also shown.}
  \label{fig:JpsiMuResults4}
\end{figure}

\begin{figure}[htbp]
  \centering 
  \includegraphics[width=0.49\textwidth]{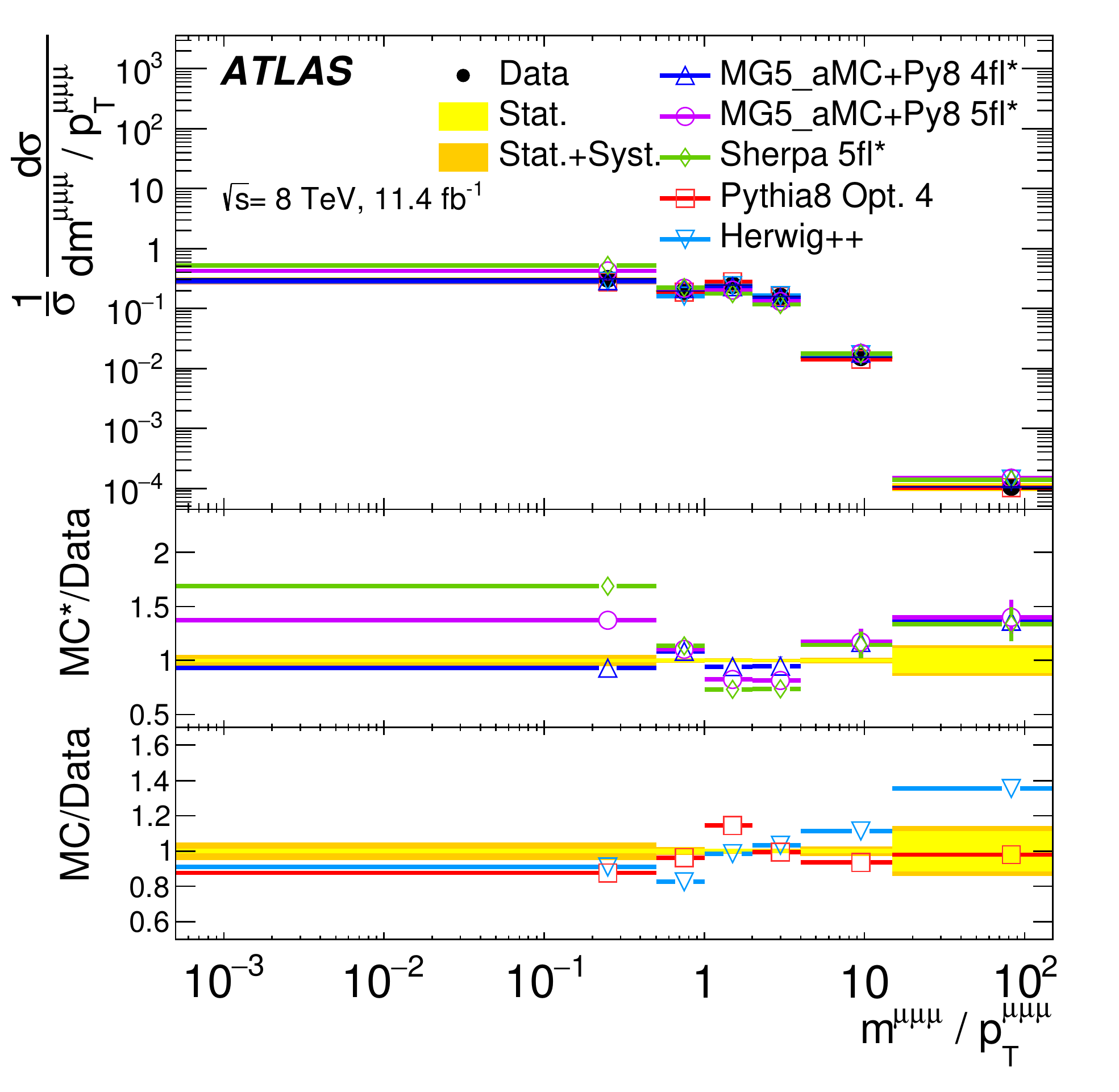}
  \includegraphics[width=0.49\textwidth]{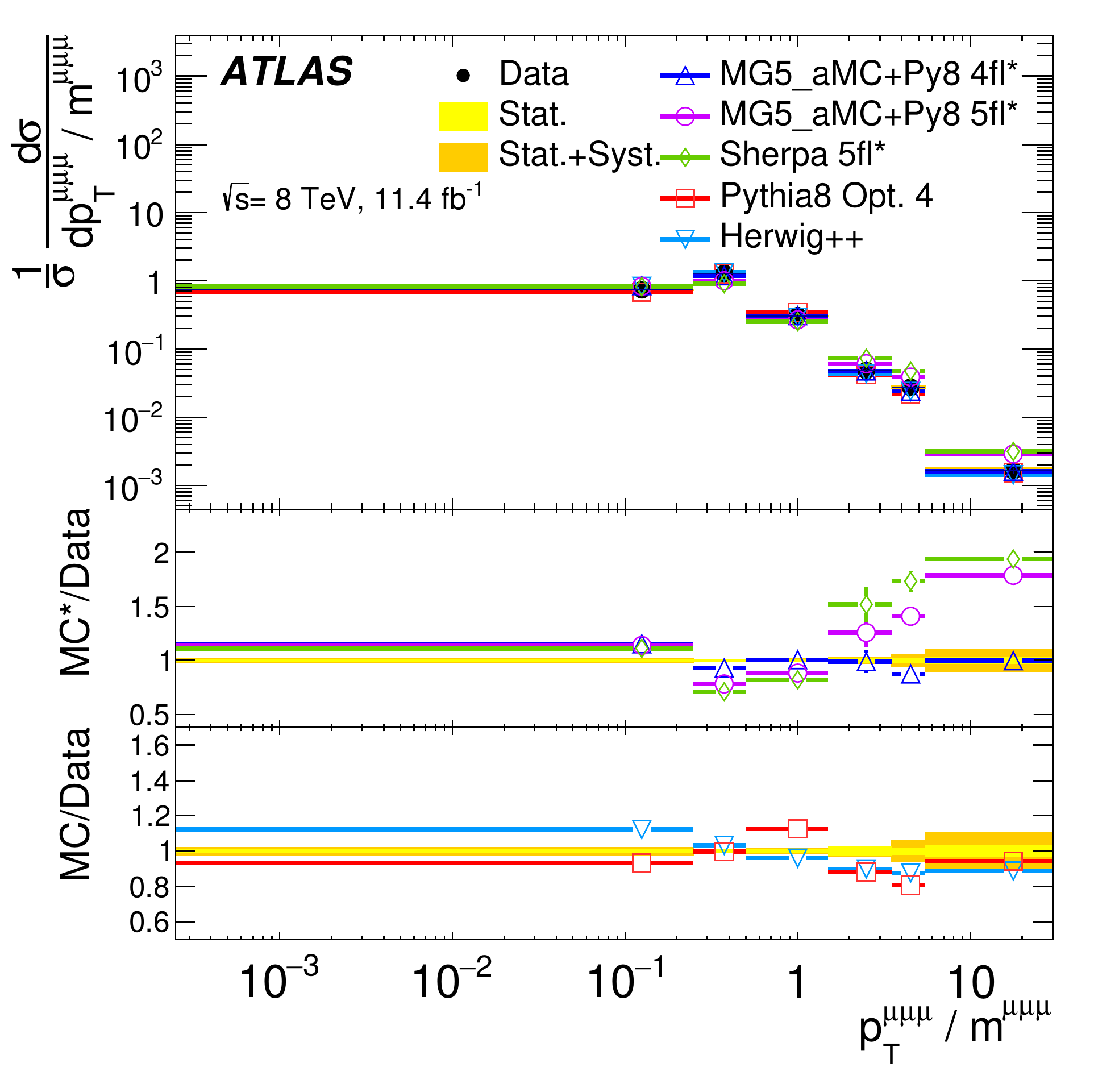}
  \caption{Measured normalised differential cross sections as a
    function of $m/\pT$\ and $\pT/m$. Comparisons are made with
    predictions of \pyegt and \hwpp. \mgamc{}+{}\pyegt and \sherpa
    predictions are also compared having been corrected from the
    two-\Bhadron production to the three-muon final state via transfer
    functions (indicated with *). The \pyegt ``Opt. 4'' gluon
    splitting parameter settings are described in
    Table~\ref{tab:py8opts}. The ratio to data of the
    \mgamc{}+{}\pyegt and \sherpa (middle pane), and \pyegt and \hwpp
    (bottom pane) are also shown.}
  \label{fig:JpsiMuResults5}
\end{figure}

Figures~\ref{fig:JpsiMuResults3}, \ref{fig:JpsiMuResults4} and
\ref{fig:JpsiMuResults5} compare the various predictions to the
measured normalised differential cross sections. The best performing
\pyegt prediction (``Opt. 4'') is included again for reference along
with the \hwpp, \mgamc{}+{}\pyegt and \sherpa predictions.
Considering first the $\DRy(\Jpsi, \mu)$ distribution, agreement
with data is slightly better for \hwpp than \pyegt. The 4- and
5-flavour \mgamc{}+{}\pyegt predictions tend to sit on either side of the
data, with the 4-flavour prediction generally being closer in
shape. The 5-flavour scheme \sherpa prediction is similar in shape to
the 5-flavour scheme \mgamc{}+{}\pyegt prediction, although the
modelling is worse overall.  The differences between 4- and 5-flavour
predictions are enhanced in the high-\pt $\DRy(\Jpsi, \mu)$
distribution, with the 4-flavour remaining close to the data while
both 5-flavour predictions move further away.

The general trends seen in $\DRy(\Jpsi, \mu)$ are also visible in
$\Delta\phi(\Jpsi, \mu)$, while in $\Delta y(\Jpsi,\mu)$ the
\mgamc{}+{}\pyegt and \sherpa predictions all provide a good
description of the data, while \pyegt and \hwpp tend to fall away at
high $\Delta y(\Jpsi,\mu)$. In $y_{\textrm{boost}}$, a distribution
expected to be more sensitive to the PDFs than to the details of event
generators, a comparable picture is seen across all predictions.

Moving to the other kinematic variables, the low $m(\Jpsi, \mu)$
region again provides discrimination between the 4- and 5-flavour
predictions, and is one case where the 5-flavour \mgamc{}+{}\pyegt
prediction lies closer to the data than the 4-flavour
prediction. However, at high values of the ratio $\pT/m$, the
4-flavour prediction clearly provides a much better description of the
data than either of the 5-flavour predictions. Indeed, considering all
distributions, the 4-flavour prediction from \mgamc{}+{}\pyegt
provides the best description of the data overall. The predictions of
\pyegt and \hwpp are generally comparable, with indications that some
further tuning could yield an improved description of the data. It
should be noted, however, that the theoretical uncertainties in these
predictions were not evaluated for this study.

\FloatBarrier

\section{Conclusion}
\label{sec:conclusion}

A measurement of the production of \Bhadron pairs in the
$\B(\rightarrow\Jpsi[\rightarrow\mumu]+X)\B(\rightarrow\mu+X)$ decay
mode has been presented, using an integrated luminosity of
11.4~fb$^{-1}$ of proton--proton collisions at $\sqrt{s} = 8$~TeV
recorded by the ATLAS detector at the LHC. A fiducial volume is
defined by requiring two muons from the decay of a \Jpsi, which itself
originates from the decay of a \Bhadron (including feed-down), and a
third muon from the decay of a different \Bhadron (including cascade
decays).  All muons are required to have \pT$>6$~GeV, the two muons
from the \Jpsi must have $|\eta|<2.3$\ and the third muon must have
$|\eta|<2.5$.  The total cross section in this fiducial volume is
measured to be $17.7 \pm 0.1 ($stat$) \pm 2.0 ($syst$)$~nb.

Normalised differential cross sections were measured for ten
kinematic observables designed to probe the underlying mechanisms
of \Bhadron production. These include a determination of
nearby \Bhadron pair production down to zero opening angle. This
region is particularly sensitive to the production of $b$-quarks via
gluon-splitting, which suffers from large theoretical
uncertainties. Constraining this region is vital for many LHC
measurements, including $H\rightarrow b \bar{b}$ in the $VH$
production mode.

Predictions for the three-muon cross section are compared
to the data. Several choices for the $g\rightarrow b\bar{b}$ splitting
kernel in \pyegt are considered, along with nominal predictions from
\hwpp, \mgamc{}+{}\pyegt and \sherpa. These cover a range of different
matrix element calculations and parton shower models, as well as 4-
and 5-flavour treatments of the calculation. Of the \pyegt options
considered, the \pt-based splitting kernel gives the best agreement
with data, performing comparably to \hwpp. The best overall agreement
with data comes from the 4-flavour \mgamc{}+{}\pyegt prediction, which
outperformed \pyegt and \hwpp, and the 5-flavour predictions from
\mgamc{}+{}\pyegt and \sherpa, though it should be noted that the
associated theoretical uncertainties must be evaluated before the
level of agreement with data can be fully quantified.

In conclusion, the measurements presented here provide a new test of QCD
calculations, motivate the choices of predictions used to model
\Bhadron production, and further tuning of parameters in the
calculations to improve the description of the data.

\section*{Acknowledgements}


We thank CERN for the very successful operation of the LHC, as well as the
support staff from our institutions without whom ATLAS could not be
operated efficiently.

We acknowledge the support of ANPCyT, Argentina; YerPhI, Armenia; ARC, Australia; BMWFW and FWF, Austria; ANAS, Azerbaijan; SSTC, Belarus; CNPq and FAPESP, Brazil; NSERC, NRC and CFI, Canada; CERN; CONICYT, Chile; CAS, MOST and NSFC, China; COLCIENCIAS, Colombia; MSMT CR, MPO CR and VSC CR, Czech Republic; DNRF and DNSRC, Denmark; IN2P3-CNRS, CEA-DSM/IRFU, France; SRNSF, Georgia; BMBF, HGF, and MPG, Germany; GSRT, Greece; RGC, Hong Kong SAR, China; ISF, I-CORE and Benoziyo Center, Israel; INFN, Italy; MEXT and JSPS, Japan; CNRST, Morocco; NWO, Netherlands; RCN, Norway; MNiSW and NCN, Poland; FCT, Portugal; MNE/IFA, Romania; MES of Russia and NRC KI, Russian Federation; JINR; MESTD, Serbia; MSSR, Slovakia; ARRS and MIZ\v{S}, Slovenia; DST/NRF, South Africa; MINECO, Spain; SRC and Wallenberg Foundation, Sweden; SERI, SNSF and Cantons of Bern and Geneva, Switzerland; MOST, Taiwan; TAEK, Turkey; STFC, United Kingdom; DOE and NSF, United States of America. In addition, individual groups and members have received support from BCKDF, the Canada Council, CANARIE, CRC, Compute Canada, FQRNT, and the Ontario Innovation Trust, Canada; EPLANET, ERC, ERDF, FP7, Horizon 2020 and Marie Sk{\l}odowska-Curie Actions, European Union; Investissements d'Avenir Labex and Idex, ANR, R{\'e}gion Auvergne and Fondation Partager le Savoir, France; DFG and AvH Foundation, Germany; Herakleitos, Thales and Aristeia programmes co-financed by EU-ESF and the Greek NSRF; BSF, GIF and Minerva, Israel; BRF, Norway; CERCA Programme Generalitat de Catalunya, Generalitat Valenciana, Spain; the Royal Society and Leverhulme Trust, United Kingdom.

The crucial computing support from all WLCG partners is acknowledged gratefully, in particular from CERN, the ATLAS Tier-1 facilities at TRIUMF (Canada), NDGF (Denmark, Norway, Sweden), CC-IN2P3 (France), KIT/GridKA (Germany), INFN-CNAF (Italy), NL-T1 (Netherlands), PIC (Spain), ASGC (Taiwan), RAL (UK) and BNL (USA), the Tier-2 facilities worldwide and large non-WLCG resource providers. Major contributors of computing resources are listed in Ref.~\cite{ATL-GEN-PUB-2016-002}.

\clearpage

\printbibliography

\clearpage
\appendix
\addcontentsline{toc}{part}{Transfer Functions}

\section{Transfer Functions}
\label{sec:transfer}
The cross sections measured in this analysis are computationally
costly to simulate using available event generators, due to the low
production rate of \Bhadrons from inclusive \MCn, and the low
branching ratio for the three-muon final state. This is potentially a
limiting factor in the use of these results for testing and tuning
theory predictions. To get around this problem, a set of
\emph{transfer functions} are derived, which can be used to translate
the predictions for a less computationally intensive inclusive
two-\Bhadron final state into the desired three-muon final
state. These transfer functions therefore capture the average
kinematics of \Bhadron decays.

The transfer functions are calculated using \MCn, by taking the ratio
of a three-muon prediction to a two-\Bhadron prediction. The
three-muon prediction is calculated for the fiducial volume defined in
Section~\ref{sec:fiducial}, and an equivalent two-\Bhadron fiducial
volume is chosen such that the shape of the $\DRy$(\Bhadron,\Bhadron)
distribution is as similar as possible to the shape of the
$\DRy(\Jpsi,\mu)$ distribution for the three-muon prediction. The
dependence of the $\DRy$(\Bhadron,\Bhadron) distribution on the
definition of the two-\Bhadron fiducial volume is studied using the
\pyegt and \hwpp event generators described in
Section~\ref{sec:result}, and based on requiring two \Bhadrons with a
minimum \pt requirement and within a defined rapidity region. The
dependence on the rapidity requirement was found to be small, so $|y|
< 2.4$ was chosen for both \Bhadrons as it lies between the $\eta$
requirements on muons in the three-muon selection. The dependence on
\pt is more significant, and the optimal value of $\pT > 15.5$~GeV was
chosen based on a fine scan of \pT selections.

While these requirements give reasonable agreement between three-muon
and two-\Bhadron selections for angular variables, the \pt\ and mass
of the two-\Bhadron system requires further treatment in order to
reproduce more closely the equivalent three-muon distribution. Due to
the additional particles produced in the decay, the muons carry only a
fraction of the momentum of the \Bhadrons, so the \Bhadron momenta are
scaled down in order to match the equivalent three-muon quantity more
closely. While no choice of value for a simple scaling parameter gives
good agreement across all \pt and mass values, a scaling factor of
1.75 provides a reasonable compromise. This scaling is only used for
mass and \pt. The two-\Bhadron fiducial volume is therefore defined by
requiring two weakly decaying \Bhadrons with \pt$>15.5$~GeV and
$|y|<2.4$. Both the mass and \pt of the combined two-\Bhadron system
are scaled down by a factor of 1.75.

Transfer functions are then derived for each of the measured
differential cross sections using the six different \pyegt
configurations described in Section~\ref{sec:result}, and \hwpp. For
each event generator, predictions for the two-\Bhadron fiducial definition
and the three-muon fiducial definition are derived independently, both
are normalised to unit area, and the ratio of three-muon to
two-\Bhadron cross section is taken. The six \pyegt and one \hwpp
transfer functions for each distribution are all slightly different,
and used to define an envelope. The geometric centre of the envelope
is taken as the default value for the transfer function in each bin,
and the spread defines an uncertainty around that central value.

\begin{figure}[h!]
    \centering
    \includegraphics[width=0.45\textwidth]{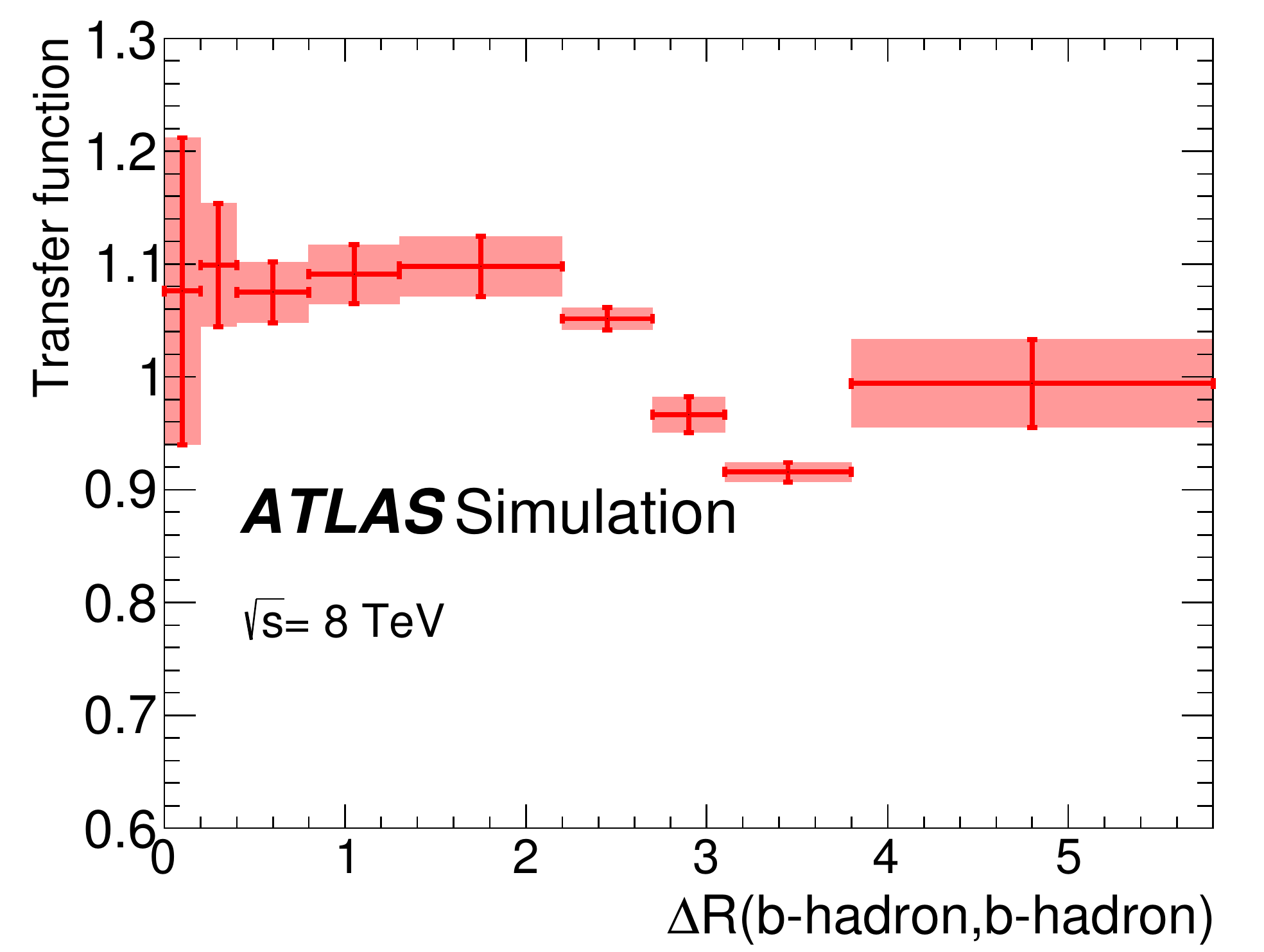}
    \includegraphics[width=0.45\textwidth]{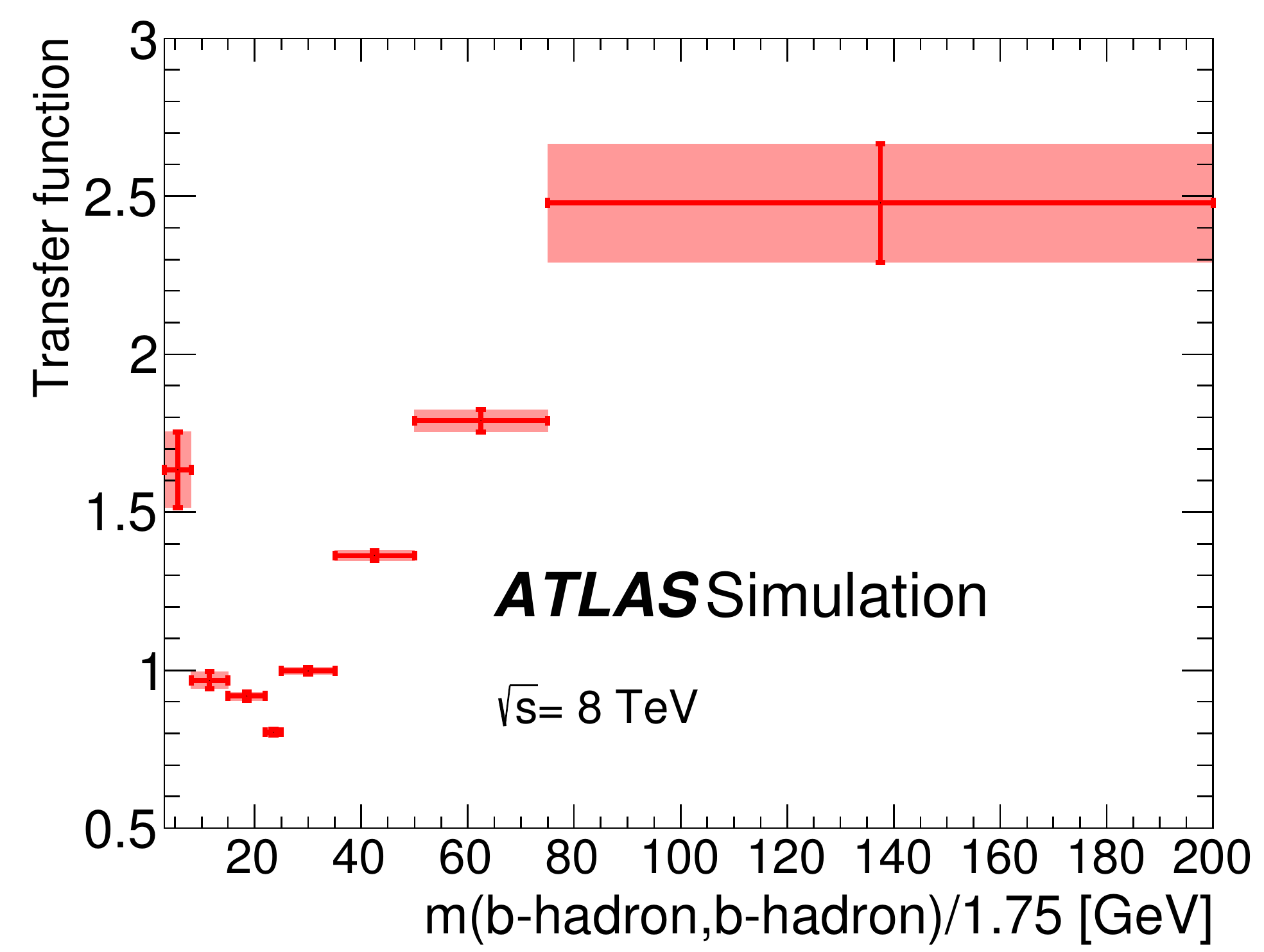}
   \caption{ The transfer functions and uncertainty envelopes for
     \DRy(\Bhadron,\Bhadron) (left) and $m$(\Bhadron,\Bhadron), with
     the 1.75 scale factor applied (right).}
    \label{fig:transfer}
\end{figure}

The resulting transfer functions for $m$(\Bhadron,\Bhadron) and
$\DRy$(\Bhadron,\Bhadron) are shown in Figure~\ref{fig:transfer}.  It
can be seen that the transfer function is relatively constant in
\DRy(\Bhadron,\Bhadron), but shows more structure in mass around the
nearby (low mass) and back-to-back (higher mass) kinematic
edges. The kinematic requirements on the \Bhadrons result in no event
populating the lowest \DRy(\Bhadron,\Bhadron) bin for the \pT$<20$~GeV
selection, and so the transfer function is not defined in this one
bin.

The transfer functions are applied as a bin-by-bin multiplicative
factor to two-\Bhadron predictions from \mgamc{}+{}\pyegt and \sherpa,
before these predictions are normalised to unit area. The uncertainty
in the transfer functions is added in quadrature to any statistical
uncertainty in the predictions from these event generators, and
completely dominates those statistical uncertainties.

\newpage 
\begin{flushleft}
{\Large The ATLAS Collaboration}

\bigskip

M.~Aaboud$^\textrm{\scriptsize 137d}$,
G.~Aad$^\textrm{\scriptsize 88}$,
B.~Abbott$^\textrm{\scriptsize 115}$,
J.~Abdallah$^\textrm{\scriptsize 8}$,
O.~Abdinov$^\textrm{\scriptsize 12}$$^{,*}$,
B.~Abeloos$^\textrm{\scriptsize 119}$,
S.H.~Abidi$^\textrm{\scriptsize 161}$,
O.S.~AbouZeid$^\textrm{\scriptsize 139}$,
N.L.~Abraham$^\textrm{\scriptsize 151}$,
H.~Abramowicz$^\textrm{\scriptsize 155}$
% ...
\bigskip
\\
% ...
$^{8}$ Department of Physics, The University of Texas at Arlington, Arlington TX, United States of America\\
% ...
$^{12}$ Institute of Physics, Azerbaijan Academy of Sciences, Baku, Azerbaijan\\
% ...
$^{88}$ CPPM, Aix-Marseille Universit{\'e} and CNRS/IN2P3, Marseille, France\\
% ...
$^{115}$ Homer L. Dodge Department of Physics and Astronomy, University of Oklahoma, Norman OK, United States of America\\
% ...
$^{119}$ LAL, Univ. Paris-Sud, CNRS/IN2P3, Universit{\'e} Paris-Saclay, Orsay, France\\
% ...
$^{137}$ $^{(a)}$ Facult{\'e} des Sciences Ain Chock, R{\'e}seau Universitaire de Physique des Hautes Energies - Universit{\'e} Hassan II, Casablanca; $^{(b)}$ Centre National de l'Energie des Sciences Techniques Nucleaires, Rabat; $^{(c)}$ Facult{\'e} des Sciences Semlalia, Universit{\'e} Cadi Ayyad, LPHEA-Marrakech; $^{(d)}$ Facult{\'e} des Sciences, Universit{\'e} Mohamed Premier and LPTPM, Oujda; $^{(e)}$ Facult{\'e} des sciences, Universit{\'e} Mohammed V, Rabat, Morocco\\
% ...
$^{139}$ Santa Cruz Institute for Particle Physics, University of California Santa Cruz, Santa Cruz CA, United States of America\\
% ...
$^{151}$ Department of Physics and Astronomy, University of Sussex, Brighton, United Kingdom\\
% ...
$^{155}$ Raymond and Beverly Sackler School of Physics and Astronomy, Tel Aviv University, Tel Aviv, Israel\\
% ...
$^{161}$ Department of Physics, University of Toronto, Toronto ON, Canada\\
% ...
$^{a}$ Also at Department of Physics, King's College London, London, United Kingdom\\
$^{b}$ Also at Institute of Physics, Azerbaijan Academy of Sciences, Baku, Azerbaijan\\
$^{c}$ Also at Novosibirsk State University, Novosibirsk, Russia\\
$^{d}$ Also at TRIUMF, Vancouver BC, Canada\\
$^{e}$ Also at Department of Physics {\&} Astronomy, University of Louisville, Louisville, KY, United States of America\\
% ...
$^{*}$ Deceased
\end{flushleft}


\end{document}